\documentstyle[11pt,aaspp4]{article}
\def\fb{$f_b$}
\def\etal{et al.~} 
\begin{document}

\title{ The Evolution of Population Gradients in Galaxy Clusters: the
Butcher-Oemler Effect and Cluster Infall}

\author{E. Ellingson\altaffilmark{1}}
\altaffiltext{1}{\footnotesize Guest observer,
Canada-France-Hawaii Telescope, operated jointly by NRC of Canada, CNRS of
France, and the University of Hawaii}

\affil{Center for Astrophysics and Space Astronomy, Univ. of Colorado, Boulder, CO, 80309}
%\email{ e.elling@casa.colorado.edu}

\author{H. Lin\altaffilmark{2}}
\altaffiltext{2}{\footnotesize Hubble Fellow}
\affil{Steward Observatory, Univ. of Arizona, Tucson, AZ, 85721}
%\email{hlin@as.arizona.edu}

\author{ H.K.C. Yee\altaffilmark{1} and R. G. Carlberg\altaffilmark{1}}
\affil{Dept. of Astronomy, Univ. of Toronto, Toronto, Ontario, M5S 3H8, Canada} 
%\email{hyee@makalu.astro.utoronto.ca,
 % carlberg@moonray.astro.utoronto.ca}

\begin{abstract}
We present photometric and spectroscopic measurements of the galaxy populations
in clusters 
from the CNOC1 sample of rich, X-ray luminous clusters at $0.18 < z < 0.55$.
A classical measure of the galaxy blue fraction for spectroscopically
confirmed cluster members shows a significant Butcher-Oemler effect for
the sample, but only when radii larger than 0.5$r_{200}$  are considered.
We perform a principal component analysis of galaxy spectra  to
divide the total cluster light into contributions from stellar populations
of different ages.
Composite radial distributions of different stellar populations
show strong gradients as a function of cluster-centric radius.
The composite population is dominated by evolved
populations in the core, and gradually changes at radii greater
than the virial radius to one which
is similar to coeval field galaxies.
We do not see evidence at any radius within the clusters for an excess of star formation
over that seen in the coeval field.  Within this redshift range, significant
evolution in the fractional population gradient is seen. Both low and high
redshift clusters have similar populations in the cluster cores, but higher
redshift clusters have steeper gradients and more star forming galaxies
at radii  outside of the core region~--
in effect, a restatement of the Butcher-Oemler
effect. Luminosity density profiles are consistent with
a scenario where this phenomenon is due to a decline over
time in the infall rate of field galaxies into clusters. Depending on
how long galaxies reside in clusters before their star formation
rates are diminished, this suggests a decrease in the infall
into clusters of a factor of $\sim 3$  between $z > 0.8$ and  $z \sim 0.5$. 
We also discuss alternative scenarios for the evolution of cluster populations.

\end{abstract}


\section{Introduction}

The evolving populations in galaxy clusters
offer a unique opportunity to directly observe galaxy
evolution, and particularly the 
effects of environment on star forming galaxies.
Present-day rich clusters have strikingly different
populations from galaxies in poorer environments,
suggesting that some mechanism is at work 
on the cluster population which is absent in the lives of 
normal field galaxies.
The first clues towards understanding the
evolution of the cluster population came more than 20 years 
ago, with the discovery that higher redshift clusters
have a larger fraction of blue galaxies than those at the
present epoch: the Butcher-Oemler effect
(Butcher \& Oemler 1978, 1984).
These purely photometric results were confirmed
spectroscopically (e.g., Dressler \& Gunn 1982, 1992; 
Lavery \& Henry 1986; Fabricant, McClintock \& Bautz 1991)
and populations of
both star forming and recently post-star formation galaxies
in moderate redshift clusters have been identified
(i.e., the Balmer-strong H-$\delta$ or K$+$A galaxies;
Dressler \& Gunn 1982; Couch \& Sharples 1987;
Poggianti \etal 1999). 
Many subsequent investigations have focused on the
details of how these galaxies are transformed into the
population seen in clusters today  (Couch \& Sharples 1987;
Barger \etal 1996; Poggianti \etal 1999; Balogh \etal 1999,
B99 hereafter).
The emerging
picture is that there may be a population of galaxies
which were formed very early in the cluster's history,
corresponding to the ellipticals often seen in cluster cores
(e.g., Bower, Lucey \& Ellis 1992; Ellis \etal  1997; 
Gladders et al. 1998;
Kelson \etal 2000; van Dokkum \etal 1998).
Subsequent generations of infalling field galaxies have
had their star formation disrupted, possibly with 
an associated starburst. As this transformation
progresses, these galaxies might be identified with normal-looking
spirals, then galaxies with strong Balmer absorption spectra,
and finally S0 galaxies which have retained some of their disk structure
but have ceased active star formation (Dressler \etal 1997).

Despite these advances, a number of questions remain about 
the evolution of cluster galaxies and the Butcher-Oemler
phenomenon. Recently, there have been several challenges to
the existence of the Butcher-Oemler effect as a universal
pattern which equally affects all classes of clusters 
(Andreon 1998; Andreon \& Ettori 1999; 
Smail \etal 1998; Jones, Smail \& Couch  2000).
The original distant cluster samples which were investigated
are a heterogeneous mixture of clusters found by optical
signatures or by their association with a strong active galactic nucleus.
In many cases, cluster masses or richnesses are
still not well determined due to the difficulty of
obtaining robust velocity dispersions, and there is a possibility
of significant substructure in some objects.
More recently, X-ray selected clusters, which in part
overlap with optically-selected clusters but on average may be
more massive or more relaxed, have been
investigated. In several cases (Smail \etal 1998;
B99), the incidence of blue or post-star forming galaxies
has been found to be significantly less than that of previous
samples. Thus, a worrisome question remains as to how
the observed galaxy populations relate to the more
global properties of the clusters, such as mass and
dynamical state, as well as the cluster sample
selection method. 

Even if clusters do contain a higher fraction of star forming
galaxies at higher redshifts, a second fundamental question
is whether this blue population represents
a true increase in the average star formation in these galaxies
over the properties of galaxies in the coeval field.
A significant fraction of galaxies in some
spectroscopic surveys have been found to be starbursts or
have Balmer lines strong enough to signify a previous
starburst encompassing 10-30\% of the galaxy mass
(Lavery \& Henry 1988; Barger \etal 1996; Poggianti \etal 1999). 
This suggests that temporary
episodes of excess star formation are associated with a galaxy's
tenure in the cluster. However, Balogh \etal (1997, B99) found
no evidence for excess star formation in a sample of
luminous X-ray selected clusters, as the generally smaller
numbers of starburst and post-starburst galaxies 
could be explained by galaxies infalling from the field.
Thus, they conclude that the cluster
environment works only in the direction of suppressing 
star formation.
The differences in these results may be partially ascribed
to differences in the  selection technique for the 
galaxies observed and identified in both cluster and field; but
some of the discrepancy may come from true differences in the
cluster samples.

Here we describe an investigation of the relationship
between galaxy evolution and cluster structure, based
on a well-defined  and homogeneous sample of
clusters at $0.2 < z < 0.6$  and coeval field
galaxy measurements. 
Photometry and spectroscopy from the 
Canadian Network for Observational Cosmology
Cluster Redshift Survey 
(CNOC1; Yee, Ellingson \& Carlberg  1996) 
were  used to construct a sample of cluster galaxies
and the properties of their stellar population.
Field galaxy measurements from the CNOC2 Field Galaxy
Survey (Yee \etal 2000) were used to accurately
determine the properties of field galaxies at these redshifts.
We present both a classical measure of the cluster
blue fraction determined from spectroscopically confirmed
cluster members and
a spectral principal component analysis (PCA)
which delineates the relative fractions of starlight from different
stellar populations.
In an approach complementary
to that of looking at individual galaxy properties in detail,
we focus on building smooth composite
spatial distributions of the various stellar 
populations. With these distributions, it is
possible to explore the relationship between the evolution of galaxy populations
and the growth and evolution of the cluster.

In \S2 of this paper we present the data and in \S3 we
discuss the blue fraction measurements.  We
describe the principal component analysis method and
tests of its accuracy from galaxies of known spectral
type in \S4. In \S5 we present the results based on composite
radial distributions of the average  PCA components.
We discuss the results and their implications
for scenarios of galaxy population evolution in rich
clusters in \S6. Our conclusions are summarized
in \S7.
Throughout this paper $q_0$=0.1 and $H_0$=100 km s$^{-1}$ Mpc$^{-1}$ are used.   

\section{The CNOC1 Cluster Sample}

The CNOC1 
Cluster Redshift Survey targeted 16 rich
X-ray luminous galaxy clusters with $0.17 < z < 0.55$.
The clusters range in velocity dispersion from about 600 to 1200 km s$^{-1}$,
with $L_X > 2 \times 10^{44}$ erg s$^{-1}$.
Deep Gunn $g$ and $r$ imaging and multi-slit spectroscopic observations
from the Canada-France-Hawaii 3.6m telescope were used to map
the cluster sample to radii of 1--3 $h^{-1}$ Mpc
from the cluster cores.
Wavelength coverage was $\sim 3500-4500$\AA ~in the rest frame
of each cluster, with a  resolution of about 15\AA.
This wavelength range samples the important features 
of [OII] emission, the 4000\AA ~break, Ca[II] H and K
and the Balmer lines H$\gamma$ and higher.
A total of $\sim$ 1200 cluster galaxies were spectroscopically identified.
The CNOC1 cluster MS0906+11 was omitted from this analysis because
it shows strong binary substructure in both X-ray and
velocity data, and robust values for dynamical parameters
were not possible.

Particular care was taken to quantify
selection effects and the completeness of the sample
as a function of galaxy magnitude, color, redshift and position.
For each galaxy, a completeness measure was empirically
calculated for its apparent magnitude, position within the
cluster and $g-r$ color ( $S_m, S_{xy} $ and $ S_{ring}, S_c$; 
see Yee \etal 1996 for details). 
These values reflect our selection
criteria for spectroscopy, as well as empirical success rates, and their
inverses are used as weighting functions to build a representative sample
of cluster properties. The largest factor in these weights is due
to apparent magnitude, and the color selection effect is very small.
The wide field coverage and careful attention to
the empirical selection functions are crucial for 
building an accurate  portrait of cluster structure.
We note that CNOC1 chose a relatively simple selection criterion
based only upon galaxy apparent magnitude and position; for comparison,
the MORPHS clusters observed by Dressler \etal (1999) were
chosen by a combination of magnitude, position and morphology/color
in order to concentrate observations on the non-passive population.
As a result, the overlap between the several clusters the CNOC
and MORPHs samples have in common is actually minimal -- MS 0016+16
has an overlap of only 7 cluster galaxies (Ellingson \etal 1998).
 
Dynamical and spatial analyses of the clusters
(Carlberg \etal 1996; Carlberg, Yee \& Ellingson 1997, CYE hereafter;
Carlberg \etal 1997a, b) yielded cluster masses, mass-to-light ratios
and density profiles.
In particular, we adopt their dynamically-determined values for 
$r_{200}$,
the radius within which the average cluster
density is 200 times the critical density. This radius averages
1.17 $h^{-1}$ Mpc for the sample;
virial radii are 1.5--2 $r_{200}$. 
Lewis \etal (1999) presented
the X-ray gas profiles from ROSAT HRI and PSPC observations
of much of the sample, finding that dynamical and X-ray
measurements of the cluster masses are in excellent agreement.
Balogh \etal (1997, B99) analyzed
spectral line indices for cluster galaxies in a classical approach to galaxy
evolution which is complementary to the approach used here.
Whereas line indices give detailed information about star
formation histories in individual galaxies, here we  use 
a PCA decomposition as a broader measure of galaxy
properties to create smooth distributions of starlight in
populations at several different ages.
These measurements are then combined with
the dynamical and X-ray properties of the clusters
to address the issue of galaxy evolution in terms 
of the spatial distribution of different populations. 

\section{The Blue Galaxy Fraction}

The first indication of evolution in cluster populations 
came from purely photometric measurements of the fraction of blue 
galaxies, \fb, in the clusters (Butcher \& Oemler 1978).
The original definition of \fb ~was based 
on the fraction of galaxies
in a cluster which are more than 0.2 mag bluer in $B-V$  than the early-type
galaxies.
The galaxies are counted to an absolute magnitude of $M_V<-18.5$ within
the radius $R_{30}$, which encompasses 30\% of the cluster galaxies.
While the Butcher-Oemler effect is clearly seen in many clusters,
this purely photometric measure can be 
affected by the significant fraction of field galaxies found
within the counting radius.
The CNOC1 survey provides a dataset from which one can 
derive \fb~for a substantial sample of spectroscopically-confirmed
cluster members, without having to perform background corrections.
Here we use similar but not identical definitions to compute \fb~for the
CNOC1 clusters, using the spectroscopically confirmed cluster
galaxies only, and for a range of absolute magnitude and radius parameters.

We define the dividing line between blue and red galaxies
relative to the observed color of the cluster red sequence. 
This guards against possible systematic errors in our color 
measurements.
On the $g-r$ vs $r$ diagram we fit the color magnitude
relation for the early-type galaxies using
a least absolute deviation (LAD) algorithm (Press \etal 1992) with
iterative rejection.
The fit is performed using galaxies inside 1/2 $r_{200}$ to maximize
the prominence of the red-sequence.
Since the LAD algorithm can be unstable, the fit is examined by eye
to ensure a good fit.
In a few instances where there are either too few galaxies
defining the red-sequence or the red-sequence has too wide a
dispersion compared to its magnitude range, an eye-fitted CMR
is used.

Blue galaxies are defined as galaxies bluer than half the $g-r$
color difference  (in the observed bands) between the E/S0 and Sbc
SED, where the SEDs are computed from the templates of  Coleman, Wu \& Weedman (1980); 
i.e., the dividing
line between blue and red galaxies is half way between the colors of an
E/S0 and Sbc galaxy.
This corresponds to $\Delta(g-r)$ of 0.21 to 0.28 mag, depending
on the cluster redshift. 
The original criterion of $\Delta(B-V)=0.2$ mag
(corrected to z=0) is similar, in that it corresponds to
$\sim$ 2/3 of the difference between the colors of an E/S0 and Sbc galaxy.
It should be noted that \fb~is strictly computed based on the
photometric properties of the galaxies, regardless of the
spectroscopic classification.
The only spectroscopic information that is used is the redshift,
which defines a galaxy as a cluster member.

For the primary \fb~values,
we count galaxies within the redshift limits defined for each
cluster as presented in Table 1 of Carlberg \etal (1996) to a k- and
evolution-corrected absolute magnitude of $M_{r}^{k,E}=-19.0$,
or the spectroscopic magnitude limit (CYE), whichever is brighter.
K-corrections are calculated from Coleman \etal (1980)  and the evolution assumed is
$\Delta M_r^{k,E} = -z$.
All but three clusters out of 15 (MS0015+16, MS0451-03, and MS1006+12)
have spectroscopic limits which are within 0.5 mag of --19.0 mag.
The counting radius is $r_{200}$.
The galaxies are  weighted
as described in \S2 above,
and error bars are computed using simple root $n$ statistics.
As a comparison to the original definition of \fb, the $M_{r}^{k,E}=-19.0$ limit is slightly
brighter (equivalent to $M_V=-18.7$), and the average radius of
1.17 h$^{-1}$ Mpc for $r_{200}$, is typically a factor of 2 or more larger
than $R_{30}$.

The results are plotted in three panels in Figure 1 as \fb~vs $z$.
Figure 1a shows the strong correlation of \fb~vs redshift, with
a range of \fb~between 0.1 to 0.4, similar to that obtained
by Butcher \& Oemler (1984).
The dashed line represents the best fit to the data, with
a slope of 0.62 $\pm $ 0.14. 
To test the effects of the galaxy sample selection limits, we also
computed \fb~values using the criteria of 
a sampling radius of $r_{200}$ and $M_{r}^{k,E}<-20.0$,
as well as
0.5$r_{200}$ and  $M_{r}^{k,E}<-19.0$
(Figures 1b and 1c).
The former test is important to verify that the somewhat uneven
absolute magnitude limits of the primary sample do not have a
significant influence on the \fb~values, as typically the higher
the redshift of the cluster, the higher the probability that it
is not sampled to --19.0 mag.
Figure 1b indicates that \fb~and the Butcher-Oemler effect do not
depend critically on 
on the luminosity of the galaxies.
Sampling to a limit one magnitude brighter produces mostly
additional scatter in the results -- the average value for \fb ~is essentially
identical with an average ratio of 1.

Sampling to a smaller radius (Figure 1c), however, appears to produce a
significant effect. 
When sampling only to a radius of 1/2 r$_{200}$,
the blue fraction drops by an average of 20\% (a 2$\sigma$ result),
and the Butcher-Oemler effect appears to be mostly obliterated, in that
most clusters (with the exception of MS0451--03) have very similar
\fb~values, regardless of their redshift. It appears that the cores
of clusters do not evolve significantly in their populations, 
and that the Butcher-Oemler effect is caused by changes in
populations at radii well outside of the cluster cores.
This result immediately hints at the possibility of strong
and evolving population gradients in galaxy clusters, and also explains some
of the discrepant results from estimates of \fb ~which have relied
on data from only the cluster cores (e.g., Smail \etal 1998).

\section{Principal Component Analysis of Galaxy Spectra}

\subsection{Technique}
Principal component analysis (PCA) provides a sensitive method
for measuring the strengths of different stellar
populations from spectroscopic data.
This technique has been used by various groups (e.g., Connolly \etal 1995;
Zaritsky, Zabludoff \& Willick 1995; Bromley \etal 1998;
Folkes \etal 1999) to determine galaxy populations.
It is especially applicable to the
faint, high redshift galaxies observed in the CNOC1 survey,
because the entire spectrum, rather than
a narrowly defined range of line indices, contributes to the measurement.

We use the relatively high signal-to-noise
galaxy spectra taken from the Las Campanas Redshift Survey 
(LCRS; Shectman \etal 1996) to derive our PCA basis vectors.
The method employed is the same as that described in detail by 
Bromley \etal (1998) in their LCRS spectral classification analysis;
here we give only a very brief summary.
The LCRS spectra are first renormalized, de-redshifted to
the rest-frame wavelength range 3500-5500\AA, and then effectively continuum
subtracted by high-pass filtering.
The resulting catalog of spectra is then put into matrix form,
upon which we apply singular value decomposition (SVD; see 
e.g., Press \etal 1992) in order to derive the set of orthonormal
vectors, here galaxy spectra, which will then form the basis in our
principal component analysis.
The original LCRS analysis finds four significant PCA basis vectors,
which correspond roughly to 
an ``Elliptical'' component showing the standard old-population features,
two ``Emission'' components exhibiting [OII] $\lambda$3727
and [OIII] $\lambda\lambda$4959,5007 emission lines typical of star formation,
and a ``Balmer'' component featuring strong Balmer absorption lines indicative of 
intermediate age stars.
In this paper we combine the two original emission components into
a single vector because our CNOC1 spectra generally 
do not extend to the H$\beta$/[OIII] region.
We plot in Figure~2 the resulting three orthonormal PCA basis vectors used in 
the analysis of our CNOC spectra.

We use a standard linear least-squares technique to decompose each 
spectrum into the best-fit linear combination of the three orthonormal
PCA basis vectors.
At the same time, we also fit a 7th-order Legendre polynomial to each 
spectrum in order to account for the continuum, as the PCA basis was
derived originally from continuum-subtracted data.
The fits are done in observed-frame wavelengths, and we
use simple linear interpolation to compute flux values for the basis
vectors at the appropriate redshifted wavelengths.
In effect, we are writing each continuum-subtracted spectrum, $s(\lambda)$,
as a linear combination of the three PCA basis vectors $v_i$,
\begin{equation}
s(\lambda) = \sum_{i=1}^{3} a_i \ v_i[\lambda / (1+z)] ,
\end{equation}
where $\lambda$ is the wavelength, $z$ is the galaxy redshift, and the $a_i$
are the best-fit PCA coefficients.
We further renormalize the $a_i$ according to
\begin{eqnarray}
c_i \equiv a_i / A , &  A \equiv \sum_{i=1}^{3} a_i,
\end{eqnarray}
so that the $c_i$ form a measure of the fractional luminosity-weighted contribution of each
PCA basis vector to the galaxy spectrum in question.
The set of $c_i$ thus provides a measurement of the stellar population for 
each galaxy, and we will typically refer to the $c_i$ as the PCA components.

The combination of these three PCA components is adequate to
separate most types of normal galaxies. Figure 3 shows PCA
components for galaxies in the Kennicutt Spectrophotometric Atlas of
low-redshift galaxies (Kennicutt 1992). Normal elliptical
and spiral galaxies are well separated on this plot.
Our chosen normalization requires that the data points 
lie within the lower left triangle of the plot: a pure
``Elliptical" spectrum will be located at the origin.
Note that the Kennicutt Atlas does not populate the region 
of the plot with large Balmer values and low emission-line values.
In this region we plot high signal-to-noise examples of K$+$A 
(post-star formation) galaxies
from the CNOC1 clusters as identified by standard line indices
(e.g., Barger \etal 1996; Poggianti \etal 1999; B99) to illustrate the
properties of these important cluster galaxies.

We estimate uncertainties in the PCA components, $c_i$, as follows.
We start with the formal uncertainties in the $c_i$ returned by the
linear least-squares fit, and check these error estimates using the
1164 CNOC2 (Yee \etal 2000) galaxies for which we have two independent spectra 
available for PCA fits.
We compare the actual differences in the $c_i$ to their formal errors, 
and we find that the main effect is that the
formal errors are underestimates at $c_i \lesssim 0.7$,
but overestimates at larger $c_i$.
We thus empirically rescale the formal uncertainties in order to match the 
observed $c_i$ differences, and as we have a large set of redundant spectral 
observations, we can readily do this rescaling
as a function of $c_i$ for each of the three PCA components.
The typical correction factors to the original formal errors 
are $\sim 1.3$ for $c_i \lesssim 0.7$, and
$\sim 0.5$ at larger $c_i$.

For our data, negative values are allowed for
any of the components, in order to retain statistical robustness
and to provide an additional check on uncertainties. 
Our average spectra have
uncertainties on the order of 0.01--0.15 in each component. 
Large negative values are absent, with the exception of a small number of
galaxies with negative ``Emission'' components. 
These stem from large noise spikes at the blue end of some of the lower 
redshift galaxies being mistaken for (negative) [OII] emission, and are usually
accompanied by large estimated uncertainties.
We note that these galaxies also have large negative [OII]
equivalent widths and large uncertainties as reported by B99.
The ``Emission'' component
is most sensitive to this type of error because we typically sample
only one strong emission feature in our spectral range; 
as such, it is most closely analogous to a single line index measurement.
Unfortunately, such errors, when combined with our normalization scheme, 
can also strongly affect the other components. 
Because of this, galaxies
with values less than $-0.2$ in any component are individually examined,
and if a single noise feature is implicated in a negative emission
component, the emission component is set to zero and the other
components renormalized accordingly. 
Comparison of galaxy colors and line indices versus PCA components,
for galaxies which have undergone this correction
versus those which have cleaner blue spectral regions, show no systematic effects.

PCA values were also checked against the classical line indices from 
B99 (Figure 4).
The expected correlations between ``Emission" and [OII] and between
``Balmer" and H$\delta$ are seen, with variations 
within the uncertainties. Note that the correlations are not
required to pass through the origin, as the PCA components 
are based on relative powers of many lines. 
The PCA analysis was able to extract usable values for
about 15\% more galaxies than the line indices, and, in general,
uncertainties were slightly smaller relative to the range
of values observed (i.e., 2\AA ~uncertainty was typical for
H-$\delta$, which has about a 10\AA ~true range in the sample;
the PCA component has a median uncertainty of 0.08 for measurements with a range
of 0--0.8). Thus, the PCA components can produce a somewhat clearer delineation
between objects with differing stellar populations.
One feature of note  is 
that the ``Emission'' component saturates at unity as the [OII] line strength
increases.
That is, our PCA technique is not usable as a linear measure of 
emission line luminosity
in individual galaxies beyond Sc, as a Scd and a starburst galaxy may both have
values close to unity.
Thus, a detailed analysis of starburst galaxies is better done directly with
emission line indices, rather than with these PCA components, though
measurements of star formation rates for less active types are still
accurate. For cluster populations, the fraction of very strong star
forming galaxies is very small, so this slight nonlinearity is unlikely
to significantly affect average population measures.

\subsection{Population Gradients in Galaxy Clusters}

A composite average population gradient
from 913 galaxies in  15 clusters was constructed
from galaxies with  k$+$ -evolution corrected $M_r^{k,E} < -19$.
In constructing composite PCA values for
subsamples of the data,
individual galaxy PCA values were weighted according to each cluster's
selection function for apparent magnitude,  position within
the cluster, and $g-r$ color (see \S2). More than 50\% of
galaxies within these magnitude limits are sampled in each cluster,
with the success rate of identifying objects observed
spectroscopically greater than $\sim 80$\%.
About 3\% of the galaxies with redshifts showed
catastrophic problems in a part of the spectrum which
made accurate PCA estimates impossible. These objects were
excluded from the sample and galaxy weights for the
remaining objects were recalculated accordingly.
Creating a fair composite sample of the galaxies as a function
of cluster-centric radius also requires
correcting for the fact that the individual clusters can have varying
observational coverage. 
To account for the fact that
some clusters were not observed to the maximum radius,
weights for galaxies at
large radii were increased by the inverse of
the fraction of the 
clusters which were observed to that radius.
For this analysis, the brightest cluster galaxy was omitted
for each cluster, since these galaxies (which are
often cD galaxies) may have a very
different star formation history from the rest of the cluster.
Finally, galaxies were also weighted by the inverse
of the cluster richness, as measured by the galaxy-cluster
spatial covariance amplitude B$_{gc}$ (e.g., Yee \&
L\'opez-Cruz 1999). This photometric measure of the
galaxy overdensity within 0.25 $h^{-1}$ Mpc of the cluster
core is well correlated with velocity dispersion and
X-ray temperature (Yee \etal 2000). Correcting for
cluster richness ensures that the results are not dominated
by the few richest clusters, and that composite  luminosity densities
for high and low redshift subsamples are properly normalized, even if there is
a difference in mean richness between them. Together, this
weighting scheme allows us to create a fair sample of
cluster properties, corrected for observational selection
effects and the gross effects of richness.
Galaxy positions were scaled by $r_{200}$ (CYE),
which corrects for
differing cluster richness by scaling each galaxy position to 
a fixed cosmological  overdensity. 
These dynamical parameters are well
determined for these clusters by 30-200 cluster members
each, and have been found to yield results in excellent agreement  with
X-ray determinations of the cluster masses (Lewis \etal 1999).

Figure 5a shows the average PCA components for a composite
of the 15  clusters, as a function of cluster-centric radius.
A clear spatial gradient is seen, with the older
population declining from the cluster core towards the outer
parts of the cluster, and an accompanying increase
in the emission line component.
There are between 107 and 294 galaxies in each bin except for
the outermost, which has only 17.
Uncertainties were estimated via bootstrap techniques.
Also plotted are the average PCA values for galaxies at
$ z \sim  0.35$ taken from the CNOC2 field galaxy
spectroscopic sample. It is clear that
the PCA component curves are approaching these values 
with increasing radius, and also that at no point is
there an excess of emission line activity in the cluster,
relative to the field, confirming the results from
Balogh \etal (1997), who used line indices from the same spectral database.

Morphological and spectral gradients similar
to this are seen in low-z clusters (e.g., Oemler 1974; Dressler 1980; Whitmore
\etal 1993).
Similar gradients, seen in galaxy color, morphology, or emission
line strengths,
have also been
noted in clusters at higher redshifts by
a number of investigators (e.g., Abraham  \etal  1996; Smail \etal 1998;
Morris \etal 1998;
Couch \etal 1998; Balogh \etal 1997 and Dressler \etal  1997).
Such gradients are
consistent with
the infall of field galaxies in hierarchical clustering
models of cluster growth (e.g., Gunn \& Gott 1972; Kauffmann 
1995; Balogh, Navarro \& Morris 2000) 
and suggest that galaxies which have more recently
fallen into the cluster potential preferentially inhabit its
outskirts.
A related phenomenon is the spatial/velocity segregation of red 
and blue galaxies in clusters. 
Blue galaxies are generally
found to have higher velocities and more extended spatial  
distributions than red galaxies, which has led to
some confusion as to the accuracy of virial mass estimates. 
Carlberg \etal (1997),
however, showed that for the CNOC1 sample, a separate dynamical analysis of red and
blue subsamples produced the same mass. Thus, we expect these gradients in our
data, and note that galaxies throughout the clusters  appear to be
at least in quasi-equilibrium with the cluster potential.

The principal component analysis allows the construction of
a smooth model cluster in the three separate components, and hence allows a
self-consistent correction for spatial projection effects.
A deprojected version of the component fractions is shown
in Figure 5b. Here an overall average galaxy density profile,
$\rho_{gal}$(r),
was adopted from CYE and is of the form:

\begin{eqnarray}
\rho_{gal}(r)= { 1 \over{r (0.66 + r/r_{200})^{3}}},
\end{eqnarray}

\noindent
where $r$ is the true, deprojected radius.
For each ring in projected radial coordinates,
we determine the fractional contribution from galaxies in shells
at greater true radius, $r$. The projected PCA
components are then corrected for contamination from the 
projection of these galaxies.
The outermost bin is the first to be corrected. 
PCA components for regions outside of the surveyed area.
The values for
$r > 5 r_{200}$ are assumed to be the CNOC2 field values and
the average of the outermost data point and the field values is assumed for
$2.5 < r/r_{200}$ $ < 5$.
The spatial deprojection is then propagated inwards, 
using the new components for the outer bins to correct the inner ones.
The spatial deprojection
tends to steepen
the gradient but the effect is relatively small; the most
significant difference is that the innermost points for the elliptical
population are raised by about 10\%.
While deprojected profiles are appealing, this process
introduces substantial random and systematic
uncertainties from the assumed profile, and errors in 
the outer bins propagate inwards~-- note, for example, how the 
slight departure from absolute smoothness in the projected 
profile at $\sim 0.6r_{200}$
is amplified in the bin at $\sim 0.4r_{200}$.
Thus, while we
provide deprojected versions of several of our results
based on the whole cluster sample, results which further split
the data into smaller subsamples are shown only in projected
spatial coordinates.

\subsection{Population Gradients in ``Natural" PCA Coordinates}

The three original PCA axes are strikingly similar to recognizable
stellar population types, but do not actually correspond exactly
to distinct real galaxy types.
However, these PCA components can be linearly  transformed to
measure a galaxy's match to any template stellar population.
In Figure 6a we plot a weighted composite for the sample, 
where
the three  components have been linearly transformed
to a new PCA coordinate system with axes named
``Old Population", ``Field-like", and ``Post-Star Formation (PSF)".
These new PCA coordinates were defined in order to divide
the stellar light from each galaxy into components representing
empirically-determined stellar populations of different ages:
a ``natural" PCA coordinate system for cluster populations.
Table I lists the axes of the redefined PCA coordinate system
in terms of the original components.
``Old Population" values were
chosen to match ellipticals in the Kennicutt Spectrophotometric Atlas
of galaxies (1992),
and agree with the reddest galaxies in these clusters.
A decomposition of spectral synthesis models (from Bruzual \& Charlot 1993)
indicates that any population
more than about 3 Gyr old will yield similar PCA values.
The new ``Field-like" vector is calculated separately for each cluster,
based on the average for CNOC2 field galaxies
at these same redshifts and  magnitudes.
On average, values are similar to a present-day Sbc, but with
additional power in Balmer absorption lines. There is a mild
increase in the ``Emission" fraction between $z=0.2$ and $z=0.6$ 
and thus the ``Field-like" component is defined as a function of
cluster redshift.
``Post-Star-Formation" values are determined 
from a spectral synthesis model of a 
galaxy whose star formation
is constant for 4 Gyr and then is abruptly stopped.
The new vector corresponds to the time of the maximum Balmer component,
about 1 Gyr after the truncation.
This rather extreme spectral type is found to be in general agreement 
with the observed PCA values from
the strongest K$+$A spectra of galaxies seen in the clusters.

Note that these components are not strictly orthonormal, though they
are normalized to a sum of unity, as before.  Thus, the equivalent of total
flux is conserved, but the final component values are dependent on the
choice of all three axes. While the ``Old Population" and
``Field-like" components are robustly determined empirically, the
``PSF" component is somewhat more arbitrarily defined. Choosing
a slightly different recipe for this component will raise or lower
all three values. For example, changing the ``PSF" component to a somewhat
less strong Balmer component, more indicative of average K$+$A
galaxies (Elliptical= 0.15, Emission=0.15, Balmer=0.70; see e.g., Figure 3) 
will raise the ``PSF" component for an average cluster galaxy by
about 0.02, and lower each of the ``Old Population" and ``Field-like" components
by about 0.01 to compensate. Thus, in addition to the random uncertainties
in our PCA values, there is an additional systematic
uncertainty of a few percent in their interpretation as real
stellar populations.

Despite this slight degeneracy,
the overall cluster structure is well illustrated by
this transformation. ``Old Population"
galaxies preferentially occupy the cluster interior, and a
smooth gradient towards younger populations is seen until
the properties of cluster galaxies 
approach those of the coeval field population
asymptotically at 2--5 $r_{200}$, well outside of the virial radius.
Figure 6b shows the same results deprojected in the same
manner as Figure 5b. The profiles are qualitatively the same
as for the projected spatial coordinates, but are steeper in the
core region. Note especially that the data are quite consistent
with zero ``Field-like" or ``PSF" component in the cluster core,
and then a fairly steep rise within about 0.5$r_{200}$.

The post-star formation component appears relatively flat in
projected spatial coordinates, and flat with a central dip in deprojected
spatial coordinates 
with a component fraction of about 10\%. Again, there is a possibility
of a few percent systematic uncertainty in the absolute values, though
the overall shape remains the same for different choices of
PCA coordinates.
This is larger than the $\sim 2\%$ fraction
of identifiable K$+$A galaxies seen in the same dataset by
B99, who further conclude that there is
no excess of such galaxies over the field.
However, the PCA technique is sensitive to
smaller fractions of galaxy light in the post-star formation state
(i.e., there is no discrete cutoff in line strength, as there
is in identifying individual K$+$A galaxies),
so higher fractions are expected than for studies which
count only galaxies with extreme signatures.
We confirm the result that there is only a few percent of
very strong Balmer-line galaxies in the clusters,
but note that there are many galaxies with slightly weaker 
Balmer lines still in excess of what is expected for normal
field or elliptical galaxies. These galaxies
overlap the observational scatter from the substantial passive/elliptical
galaxy population and so individual identification of their populations
is impossible.
Thus, the comparison with B99 appears reasonable, and 
this analysis of the data suggests that there is a small excess
of light from intermediate age stars over the field population.
Comparison with the K$+$A fraction in the 
MORPHS sample of clusters at similar
redshifts (Dressler \etal 1999; Poggianti \etal 1999) 
is slightly more troubling, as they find closer
to 20\% of the galaxies have strong post-starburst signatures. 
This  may signify true differences in cluster populations
for X-ray selected clusters such as the EMSS sample,
versus the optically-selected MORPHS clusters (see \S6). 

The intermediate shape of the ``PSF" profile, neither rising nor falling as steeply
as the other components, indicates that the ``PSF" components
do indeed trace a population which is intermediate in 
dynamical state as well as stellar age.
Interestingly, because the populations must approach a pure
``Field-like" component outside of the cluster,  we expect 
that the ``PSF" component 
will approach zero at large radii. However, our data do not show
an obvious decline even in the outermost data point.

\subsection{ The Evolution of Population Gradients}

In Figure 7 we examine the trend in population gradients in clusters
over our redshift range of 0.18 to 0.55.
Plotted are the ``Old Population"
components versus radius for subsamples at $0.18 < z < 0.3$ and
$0.3 < z < 0.55$. The PCA values,
weights and uncertainties are
calculated as before. While the cluster cores
appear to be similarly dominated by old populations,
a significant change in the fractional population gradient is seen 
between the two redshift bins. At lower redshift, the old
population appears to dominate to larger radii, whereas
at higher redshifts, the field-like galaxies are more noticeable
in the inner portions of the cluster, though they still avoid
the central core. 
This steepening of the population gradient
can be thought of as a more detailed restatement of the
Butcher-Oemler effect. Note that redefining the
Butcher-Oemler effect in terms of gradients immediately
explains the changes in the apparent blue fraction measures
shown in Figures 1a and 1c, where the counting radius 
was changed. It also explains why some studies which have
focused on the cluster cores  have found
very little evidence for population evolution.

A rough comparison of these results with those from previous
measures of galaxy cluster populations can be made by
equating our ``Old Population" with the summed fractions
of E and S0 galaxies from morphological studies and
assuming that $r_{200} \sim 1$ $h^{-1}$ Mpc for clusters
in other surveys. 
Values inferred from Figure 7 of Dressler \etal (1997)
for the subset of ``regular" clusters from the MORPHS sample 
are plotted with a dashed line on Figure 7, and suggest that the PCA population gradient
in the higher
redshift bin are generally
in agreement with other clusters at similar redshifts.
Interestingly, our lower redshift bin, with a mean
redshift of $\sim 0.23$, appears to be very close to
the values inferred by the $z \sim 0$ morphological
gradients derived by Whitmore, Gilmore \& Jones  (1993).
The dotted line on Figure 7 shows an estimate of the
population gradients from their low redshift sample.
While care must be taken in comparing spectral and morphological
quantities, it appears that the PCA gradients 
are generally in agreement with other measures of galaxy type,
and also suggest that the evolution in the gradients tapers off
at $z < 0.2$. 

\section{Luminosity Density Profiles}

Luminosity 
surface density profiles for each PCA component were constructed  from
the total galaxy luminosity surface density distribution
and the PCA fractional gradients shown in Figure 6a. 
Figure 8 
shows the profiles for the transformed PCA coordinates
described above. 
The fractional gradients seen in Figure
6 are shown here as relative light distributions
which increase in spatial extent from older to younger stellar populations.
This is quite consistent with what is expected from
the accretion of field galaxies into the cluster,
where galaxies which fell into the cluster at earlier times
and whose stellar populations have longest felt the
effects of the cluster environment  will
have a more concentrated spatial distribution than those that are
only now entering the cluster for the first time.
We note that the ``Field-like" and ``PSF" populations are most
similar in shape, suggesting that much of the older 
population significantly predates the introduction of
these younger galaxies to the cluster. 

In Figure 9 we illustrate a possible explanation for the
evolution in cluster populations and population gradients
in clusters by constructing luminosity profiles
as a function of cluster redshift.
Figure 9a shows the relative
luminosity surface density for the ``Old Population" component for
$0.18 < z < 0.3$ and $0.3 < z < 0.55$.
The two curves are nearly identical in both shape and
normalization.
Note that the normalization for all curves
in Figures  8, 9a and 9b is identical~-- since
the weights for
each galaxy are 
normalized by the cluster richness, as measured by B$_{gc}$, the possible observational selection
effect that the higher redshift bins contain somewhat richer clusters on average is removed.
Plotting the distributions as a function of $r/r_{200}$
rather than a fixed metric radius normalizes each galaxy
position in terms of cosmological overdensity at the cluster redshift.
It appears that the old population, which presumably includes both 
galaxies which were formed early in the cluster's history, as well
as galaxies which have been spectroscopically transformed after
falling into the cluster environment,
has not changed in shape significantly relative to $r_{200}$ over this timescale,
This is in agreement with the idea that these galaxies have resided in the
cluster for several billion years and are in
a stable dynamical equilibrium. 

Figure 9b shows the sum of the ``Field-like" and ``Post-Star-Formation" 
light profiles, representing galaxies which are newer introductions
to the cluster potential. 
The two components were summed in part to slightly increase the signal-to-noise
ratio and also to represent more closely the light from 
``Butcher-Oemler" galaxies, which are often  taken as the mixture of both star
forming and K$+$A galaxies. Summing two of the three components
also minimizes possible complications from the
PCA coordinate transformations,
since the cluster populations are now divided into two nearly-normal
components, with a systematic uncertainty from PCA coordinate
definitions of only about 1\%.
These curves are normalized identically
to those in Figure 9a; note that the core 
densities are lower by about a factor of 8 
and the curves 
are more extended, relative to
the ``Old Population."
Also, unlike the ``Old Population", the young
stellar populations show significant evolution
over the redshifts observed. Qualitatively, the two
curves appear to have similar shapes outside of the cluster core,
but the lower redshift curve has an amplitude about
a factor of three lower than the higher redshift curve.
Thus the Butcher-Oemler effect appears to be caused
by a decrease in the density of a young stellar population
which has the same overall spatial extent relative to
$r/r_{200}$ at these redshifts.

\section{Discussion}

\subsection{The Infall of Field Galaxies into Clusters}

The results from composite radial distributions of
galaxy populations are consistent with a simple model
of cluster formation and evolution. It appears that
the cluster can be  modeled as
the sum of two components: a
virilized component of older galaxies,
and a younger component  which, 
while it
is probably in quasi-equilibrium with the cluster potential
(Carlberg \etal 1997a), has fallen in more recently and which 
may eventually be expected to transform both spectrally 
and dynamically to blend with the older population.
Galaxies in the
midst of this transformation (e.g., K$+$A galaxies,
or the ``PSF" PCA component) appear to inhabit
an intermediate spatial distribution.

The origin of the young and intermediate-age components -- identified with the
``Butcher-Oemler galaxies" -- may be constrained by tracing
their spatial luminosity distribution as a function of redshift.
Over the redshift range sampled here,
we find that outside the core region ($>$ 0.25 $r_{200}$)
the shape of the young population luminosity distribution appears to be unchanging.
This strongly suggests that dynamical state of galaxies (i.e.,
time since entry to the cluster potential) and age of their
stellar populations are fundamentally linked. This is a reasonable
assumption if the mechanism which transformed the population
is due to the cluster environment.
The main difference in these curves is then a simple vertical
shift, and these results can thus
be interpreted as a straightforward decline in the infall rate of
new galaxies into the cluster.  In this scenario, galaxy 
population changes can be used as a direct measure of
cluster growth rates.
This explanation for galaxy population evolution was also suggested by
Kauffmann (1995), who modeled cluster
populations according to a recipe for hierarchical clusters
in a CDM universe, and Abraham \etal (1996), who modeled
the morphological gradients in the CNOC1 cluster Abell 2390
as a function of cluster infall.

The young, field-like component comprises
about 15\% of the galaxy light within $r_{200}$ for our
higher redshift subsample.
The difference in mass-to-light ratios between the ``Field-like" and 
``Old Population" stars 
can be estimated from the relative flux at $\sim 4000$\AA~ 
in population synthesis models for a galaxy with ongoing
star formation versus a passively evolving population
(e.g., Charlot \& Bruzual 1993).
The mass-to-light ratio of the ``Old Population" should
be about 2.5 times that of the ``Field-like" population,
implying that about 6\% of the stellar mass in the cluster
has recently fallen into the cluster at an observed
epoch of $z \sim 0.45$. 
Thus, the Butcher-Oemler effect illustrates the sensitivity
of galaxy populations in tracing even a very small fraction
of infall into the cluster.

Figure 9 further suggests that the infall rate
has declined further by a factor of $\sim 3$ by $z \sim 0.2$.
However, calibrating this result to absolute
infall rates and linking it to cosmological models remain
uncertain because of the unknown timescales of both virialization
and population evolution in these environments. For example, 
the mechanism which affects ongoing star formation in clusters
might operate quite gradually, as may be indicated by
the relatively weak Balmer line strengths. If this is the case, 
the ``Field-like" galaxies we see in
the cluster may have been introduced to the cluster a billion
or more years earlier than the observed epoch. That the
blue galaxies are seen to be in equilibrium with the same
mass potential as the red galaxies (Carlberg \etal 1997a) also suggests that they may
remain in the cluster for several crossing times before
undergoing a significant spectral transformation. In addition, the spectral
transformation can be expected to take 1--3 billion
years, even if star formation is terminated abruptly.

The ratio of the ``Field-like" to ``PSF" components in Figure 8 places
rough limits on the maximum length of the delay between
galaxy infall and the cessation of star formation. The observed ratio
in the light densities of about 4 implies a  ratio of  2--3 
in the mass densities, since the mass-to-light ratio of the ``Field-like"
population will be less than that of the more evolved ``PSF" population.
The ``PSF" population is relatively short lived, with a duration of
about 1 Gyr.
Thus, in the case of a constant infall rate, the ``Field-like" population
can endure the cluster environment for 2-3 Gyr.
For a declining
infall rate, this value is an upper limit.
Estimates of this timescale were also found to be on the
order of a few Gyr by Poggianti \etal (1999) and Balogh \etal (2000).
The observed
epochs of our redshift subsamples are at $z \sim 0.23$ and $ z \sim 0.43$;
a conservative estimate of a 1.5 billion year delay
between infall and the 
post-star formation phase  of the  spectral transformation 
would place the epoch of infall corresponding to the observations
at $z_{infall} \sim 0.5$ and $\sim 0.85$,
for the lower and higher redshift bins,
respectively 
($H_0$= 75 km s$^{-1}$ Mpc$^{-1}$, $q_{0}$=0.5).
Note that this assumes that star formation
is terminated almost instantaneously upon a galaxy's entry into the cluster. 
A larger delay of 3 Gyr places the infall 
for our larger observed redshift bin  at $z_{infall} = 1.6$.

The {\it very} tentative agreement between our $ z \sim 0.2$
spectroscopic gradient and the morphological gradients
from $ z\sim 0$ (Whitmore \etal 1993)  also suggest that 
cluster infall is very rapidly decreasing during intermediate redshifts 
and perhaps is coming to a steady state by our lower redshift bin,
corresponding to $z_{infall} \sim 0.5$.
This is qualitatively in agreement with evidence that the cluster
X-ray luminosity function has not evolved strongly since
$z \sim 0.8$ ( e.g., Henry \etal 992; Rosati \etal 1998; 
Vikhlinin \etal 1998), and low-density cosmological models which place
most of cluster formation at $z > 1$. 
Expanding the study of population gradients to both higher and
lower redshifts, and combining the results with analytic
and N-body results for cluster growth will be a topic of
further investigation.

\subsection{Other Scenarios for Galaxy Evolution in Clusters}

An alternate viewpoint to a declining infall rate is that
some physical  property of the cluster environs has
evolved over these timescales, driving the observed evolution
in the populations. Possibilities include 
differing rates or efficiencies of galaxy-galaxy interactions,
evolving gas densities in the cluster, which would affect the efficiency
of ram-pressure stripping of gas from infalling galaxies, and the
evolution of the infalling field population (Lavery \& Henry 1988;
Dressler \& Gunn 1993; Moore et al. 1996; Fujita 1998).
We have already removed the primary  effect
of field galaxy evolution by constructing vector components
which are based on the observed field population at the
same magnitudes and redshifts as the clusters.

Global evolution in the hot intra-cluster medium is also probably not a strong
driving force in the observed evolution. All of the clusters
in our sample are luminous X-ray clusters, and there does
not appear to be a strong correlation between the X-ray
gas density and the galaxy populations. 
Figure 10  shows
the ``Old Population" component for 8 clusters versus
azimuthally averaged X-ray gas density, each measured at 
0.5 $r_{200}$ from the cluster core. The gas density
measurements are based on the ROSAT/ASCA observations from
Lewis \etal (1999). 
While there is some
variation in each parameter, they do not appear positively
correlated, as would be expected if gas density evolution
were driving the overall population changes. (Note that gas stripping
may still be responsible for the transformation of individual galaxies,
while an independent, cosmologically changing condition drives the population
evolution.) The baryon fractions in these clusters also do
not show a strong correlation with redshift (Lewis \etal 1999),
further suggesting that the Butcher-Oemler effect is not directly
linked to an evolving intra-cluster medium. 

Changing rates of efficiencies of galaxy-galaxy interactions 
remain a possible mechanism for
galaxy evolution in clusters, although morphological evidence
for this remains mixed (e.g., Oemler \etal 1997). Since
the effects of cluster richness and dynamics are normalized
in this analysis, any evolutionary effect must come from either
a cosmological change in the clustering of galaxies as they
fall into the cluster (i.e., a declining rate of
infalling pairs or small groups which might 
inhabit the regions near clusters)
or in the ability of galaxies to continue to form stars in the
wake of a collision. A more detailed analysis of
statistical interaction rates in these clusters is underway.

A common property of these scenarios is that either
more favorable environmental conditions or a higher resiliency of the
infalling field galaxies result in star formation in infalling galaxies
persisting  for a longer time within the high redshift cluster
environment. This general scenario of a longer
interval before a galaxy completes its spectral
transformation can be contrasted against the
alternative scenario of changing infall rates, and
the observed trend in the fractional population gradients
can be used to constrain them.
As discussed above, the observed gradients are 
qualitatively in agreement with the scenario that
changing infall rates drive the observed evolution.
A key feature of the infall model is that most of the
additional blue galaxies seen at higher redshift are linked
with a newly infalling population with an extended spatial
distribution. Thus the Butcher-Oemler effect is 
greater when a larger counting radius is used and the
cluster core is changing relatively slowly.

In the changing-timescale scenario, the excess blue
galaxies seen at higher redshift are galaxies which 
are persisting in their star formation despite the fact
that they fell into the cluster long ago.
Thus 
we would expect to see that the  major evidence for
evolution would be that the ``Field-like" population
would have a more concentrated spatial distribution 
relative to $r/r_{200}$
(i.e., more like the ``Old Population")
at higher redshifts.
In  terms of population
gradients, this will tend to create gradients which
manifest the Butcher-Oemler effect via additional blue galaxies
in the inner parts of the clusters.  Thus, the gradients would tend to
become flatter at higher redshift, with the strongest evolution
in the core region due to the
dynamically old but spectroscopically
unevolved population. 
Figures 7 and  9 cannot definitively rule out this possibility,
as the strength of such an effect is based
on the details of both cluster and galaxy formation, as well
as the exact mechanism that truncates star formation in the cluster
environment.
However,
the data appear to be more consistent with a simple infall
model rather than with a simple model of increasing the length
of time galaxies can persist in forming stars in the cluster
environment.

\subsection{How Representative are Current Cluster Samples?}

While some of the discrepancies in measurements of
the Butcher-Oemler effect can be explained by differing
observational and analysis techniques, there appear to be
significant differences in the average properties of different
cluster samples. In particular, the CNOC1
X-ray selected sample appears to have a significantly lower
fraction of both star forming and post-star forming galaxies
than the optically-selected MORPHS sample  (B99;
this work).  This difference most likely springs from 
the two different ways in which the samples were collected.
The MORPHS sample for many years  comprised the only sample of
rich clusters at $z > 0.2$ and is a heterogeneous collection
of optically and AGN-selected clusters.
They span a fairly wide range of richnesses, and in some
cases the richness or cluster mass is still not well-determined;
overestimating the mass because of velocity structures in
a merging cluster may be a concern. 
Morphologically they also span a range of types, including
both centrally concentrated and diffuse clusters and it is not
clear how well they represent the cluster population as a whole.

The CNOC1 cluster samples were chosen primarily from the
Einstein Medium Deep Survey (EMSS; Gioia \etal 1990, Gioia \& Luppino 1994) 
X-ray sample,
with the addition of the rich cluster Abell 2390. 
Both optical and X-ray morphologies show that these clusters
are on average very smooth and regular. This is not surprising, as
the  X-ray selection technique
is most sensitive to the high gas densities in the cores
of centrally concentrated  clusters. In addition, the central luminosity peaks 
from cooling flows can make less-massive high redshift clusters
much more visible to X-ray surveys: nearly half of the CNOC1 clusters
are  cooling flow clusters. As cooling flows may be associated
with clusters which have {\it not} had a major merging event
in the recent past (e.g., Allen 1998), this may strengthen the tendency towards
regular clusters. Hence, selection technique may predispose  the
CNOC1 sample, and very luminous X-ray clusters in general, to appear to
be more quiescent than other samples, and our implied  infall
rates may actually be lower limits to the actual growth of cluster
structures. 

These systematic differences in the cluster samples may be
responsible for the differing fraction of post-starburst galaxies
seen in the MORPHS and CNOC1 samples, if the starburst phenomenon
is associated with the infall of small groups into the
cluster potential. Starbursts may be commonly induced by
galaxy interactions, which in turn are more frequently
found in the low velocity dispersion environments of small groups.
If the luminous X-ray clusters which dominate the CNOC/EMSS sample have
not absorbed a significant subclump of galaxies within the
previous 1.5 or more Gyr, they are not likely to contain
a significant fraction of galaxies with post-starburst spectra.
A number of new surveys for high redshift clusters are amassing
large numbers of both fainter X-ray clusters and optically-selected
clusters (e.g., Zaritsky \etal 1997; Holden \etal 1999; Scharf \etal 1996;
Rosati \etal 1998; Kim \etal 1999;
Lewis, Stocke \& Ellingson 1999; 
Gladders \& Yee 2000).
These should soon allow 
galaxy population studies for complete samples of clusters
which are truly representative of the cluster population at
a chosen redshift, and provide enough variation in cluster
type to investigate the dependence of population on the clusters' 
recent merging histories.

\section{Summary and Conclusions}

We have  investigated the Butcher-Oemler effect in a
sample of 15 rich galaxy clusters at $0.18 < z < 0.55$
from the CNOC1 cluster survey.
A classical measurement
of the cluster blue fraction, based on spectroscopically
confirmed cluster members,
shows a significant Butcher-Oemler effect, in that
high redshift clusters have a larger fraction of
star forming galaxies. These galaxies are
found primarily outside of the cluster cores, and the
Butcher-Oemler effect virtually disappears when
a counting radius of only 0.5$r_{200}$ is used.

We have developed an alternative method of measuring
stellar populations in galaxies via a principal component
analysis (PCA) of galaxy spectra. Basis vectors defined 
from the low-redshift LCRS spectral database were used to decompose
cluster galaxy spectra into three components, corresponding
to an old, passively evolving population, a young population
matched to the coeval field, and a population of intermediate
age, matched to a 0.5 Gyr post-star formation population.
These components are shown to be robust measurements of
stellar populations by comparison with low redshift galaxy spectra
and classical line indices. However, unlike line indices,
PCA components directly measure the fraction of light emitted by
a given population, allowing for straightforward construction of
population fractions and spatial distributions.

Composite population
gradients show a smooth transition from the
infalling field galaxy population to the older populations
seen in the inner regions. 
The gradients are corrected for the evolution of the field
population and cluster richness and show no evidence
for an excess of star formation in the cluster; the
cluster environment therefore appears to have the
monotonic effect of decreasing star formation in
infalling galaxies. However, there is evidence for
a small excess of intermediate age stars, mostly
located in galaxies with mixed stellar populations.
This suggests that star formation in these galaxies has 
been truncated gradually, without an accompanying strong starburst.

Evolution in the cluster galaxy population between $z=0.5$ and 0.2
is manifest in a flattening of the fractional population gradient at later epochs,
while the cluster cores remain dominated by an old, red
population. 
Luminosity surface brightness distributions indicate
that the old stellar population has a centrally concentrated, unchanging
spatial extent as a function of $r/r_{200}$ . Younger galaxy populations have
a more extended distribution that also retains its general shape
but appears to be decreasing
by about a factor of three in amplitude between $z=0.5$ and $z=0.2$.
This phenomenon is most consistent with scenarios where
the mechanism which truncates star formation in individual galaxies
remains constant, but the cluster population evolution is
driven by a declining rate of infall into the
clusters. 
Depending on the length of time that star formation in galaxies can
survive the cluster environment, this may indicate a 
change in the infall rates into clusters between $z > 0.8$ and
$z \sim 0.5$; if galaxies reside in the cluster
for a significant interval before ceasing star formation,
the evolution will correspond to even earlier epochs.
We note that the stellar mass fraction of young populations
in these clusters is on the order of a only few percent, despite the
fact that it requires several Gyr to transform stellar populations
from young to old populations.   
This also suggests that the Butcher-Oemler
effect as seen at intermediate redshifts is a lingering 
echo of more rapid cluster formation which occurred much earlier,
in qualitative  agreement with low-density cosmological models.
Additional N-body and analytical modeling is necessary to
firmly link the observed evolution in population gradients to
models of cluster formation and growth. 

These results indicate that galaxy populations provide
a very sensitive probe of cluster growth, and that  the
construction of spatial distributions for different
galaxy populations may illuminate both cluster
and galaxy evolution. Extending similar studies to both
higher and lower redshifts is critical towards building a
consistent picture of cluster evolution.
However, the best studied
samples at intermediate and high redshifts may not be
representative of the cluster population, as both
X-ray and optically selected samples may introduce systematic
biases  in the dynamical state and formation history
of the clusters.
As the populations seen in individual clusters
may be strongly affected by the recent merging history
of the cluster, it is important that a wide variety of
cluster types be sampled in order to build a truly
representative sample.
A new generation of high redshift cluster samples based on
algorithms which are sensitive to low surface brightness 
and newly merging cluster structures is necessary to expand and
clarify our picture of
the formation of galaxy clusters and the effects of environment on 
galaxy evolution.

\acknowledgments
EE acknowledges support provided by the National Science Foundation
grant AST 9617145.
HL acknowledges support provided by NASA through Hubble Fellowship grant
  \#HF-01110.01-98A awarded by the Space Telescope Science Institute, which 
  is operated by the Association of Universities for Research in Astronomy, 
  Inc., for NASA under contract NAS 5-26555.
The CNOC surveys are supported by operating and collaborative grants from NSERC 
to HY and RC.

\clearpage

%Insert Table I here
\begin{table}
\caption{{\bf ``Natural" PCA Component Axes}}
%\scriptsize
\begin{tabular}{lcccc}
\tableline \tableline
New component Axis & & Elliptical  & Emission  & Balmer \\ 
\tableline
Old Population & & 0.77 & 0.03 & 0.20 \\
Field-like & $z$=0.25 &  0.32 &    0.39 &    0.29   \\
 & $z=$0.35 &  0.28  & 0.46 &  0.26   \\
 & $z=$0.45 &  0.22  & 0.51 &  0.26   \\
 & $z=$0.55 &  0.22  & 0.52 &  0.26   \\
Post Star Formation (PSF) & & 0.10  &0.05  &0.85   \\
\tableline
\end{tabular}
\end{table}

\clearpage

\begin{figure}
\plotone{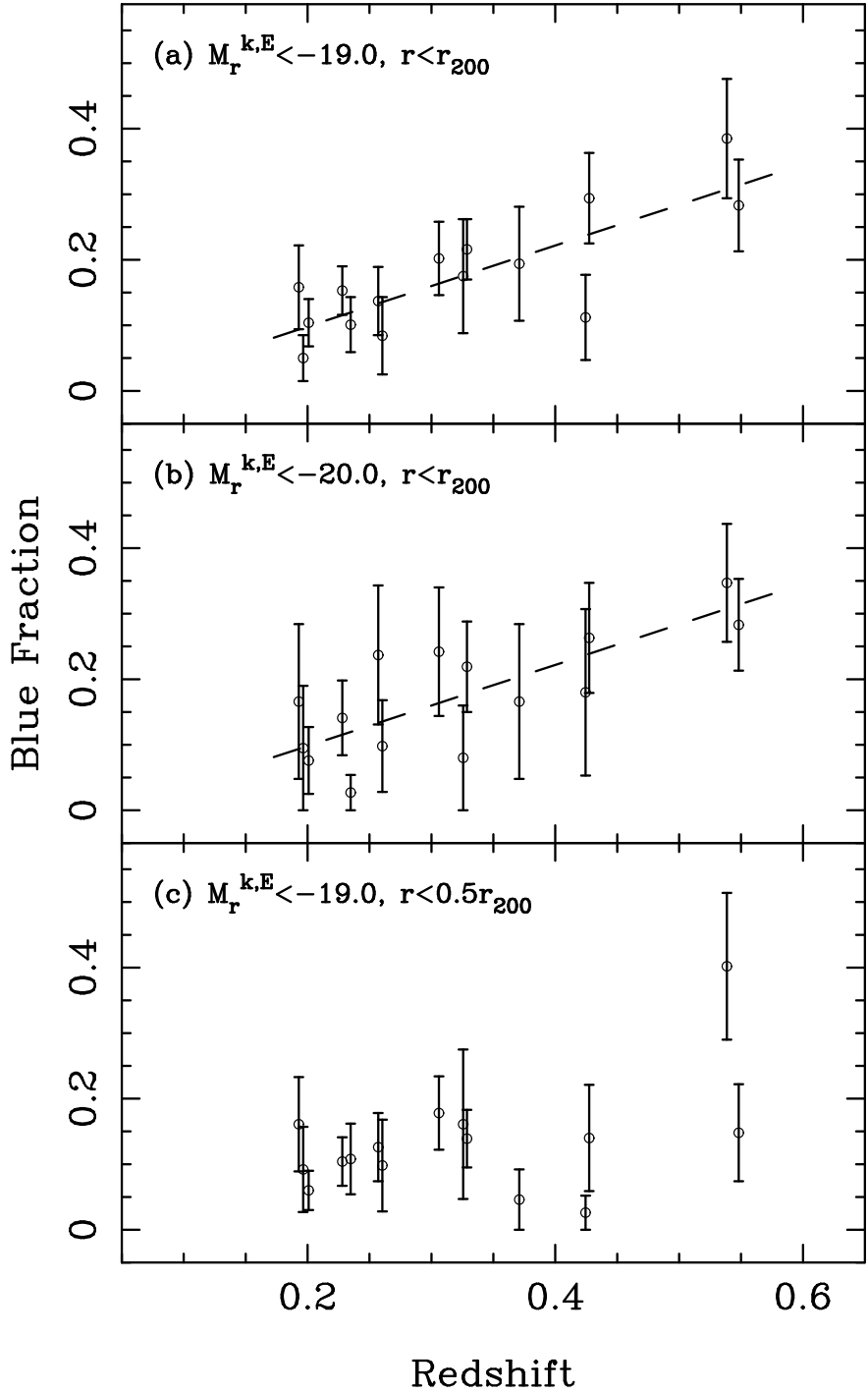}
%\plotfiddle{bo.ps}{4.0truein}{0.0}{50}{50}{-200}{0}
\caption{Blue fraction for spectroscopically
confirmed cluster members versus redshift. 
The data shown in panel a) uses a counting radius of $r_{200}$ and
a k$+$e corrected absolute magnitude limit of $M_{r}^{k,E}=-19.0$.
Values for \fb ~in panel b) are based on a magnitude limit of
--20.0, and in c), a radius of 0.5$r_{200}$ and 
a magnitude limit of --19 were used.
The dashed line in panel a) is the best fit to the data.
The same line is plotted in panel b) to illustrate 
that a sample with a brighter limiting magnitude also shows the 
Butcher-Oemler effect, but with a larger scatter.
}
\end{figure}

\begin{figure}
\plotone{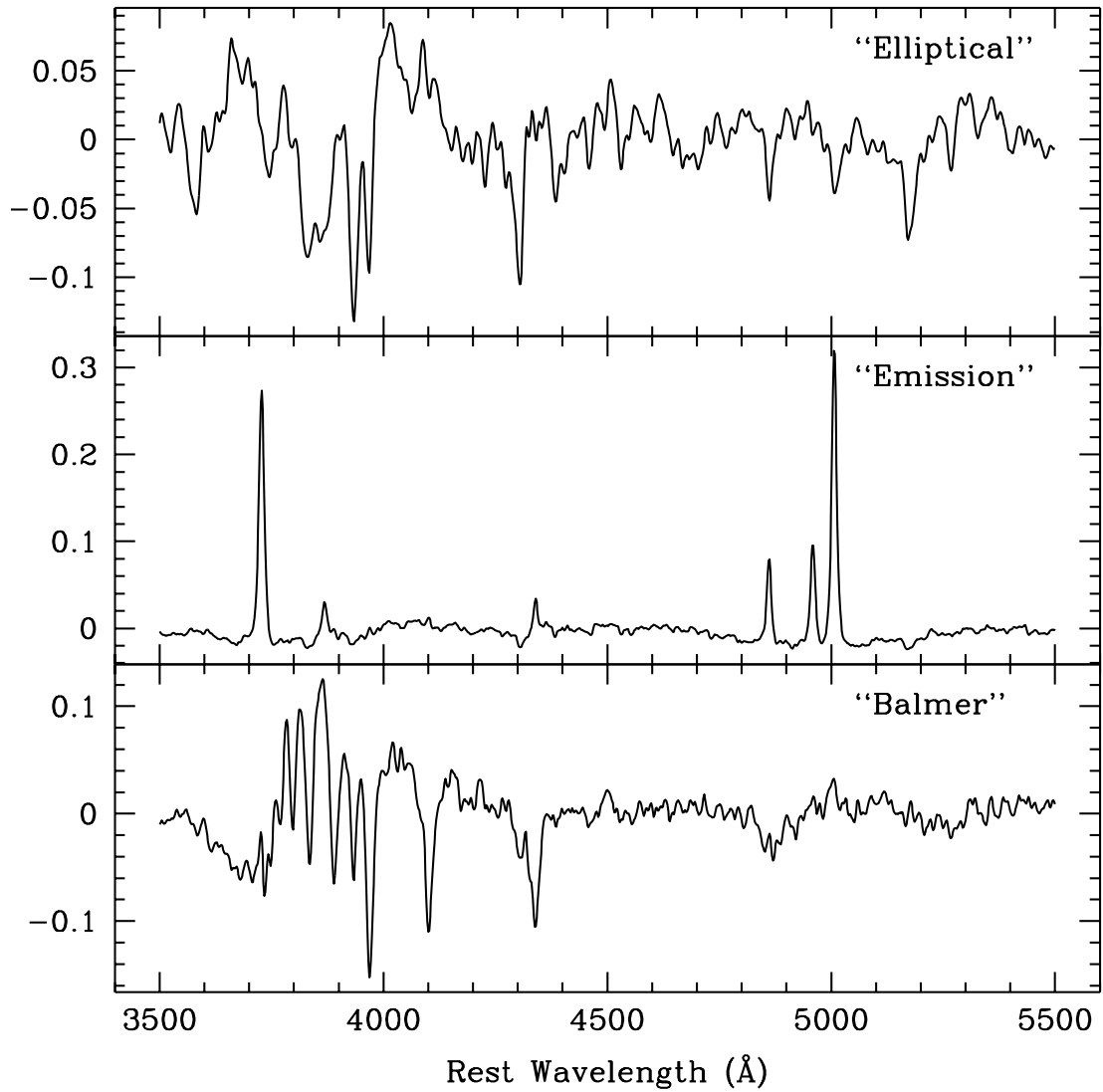}
\caption{Principal Component Vectors derived from the Las 
Campanas Redshift Survey. Three vectors are used: 
``Elliptical", ``Balmer", and ``Emission" (combined from
two original vectors; see text).
}
\end{figure}

\begin{figure}
%\plotfiddle{kenn.ps}{1.5truein}{0.0}{50}{50}{-140}{-10}
\plotone{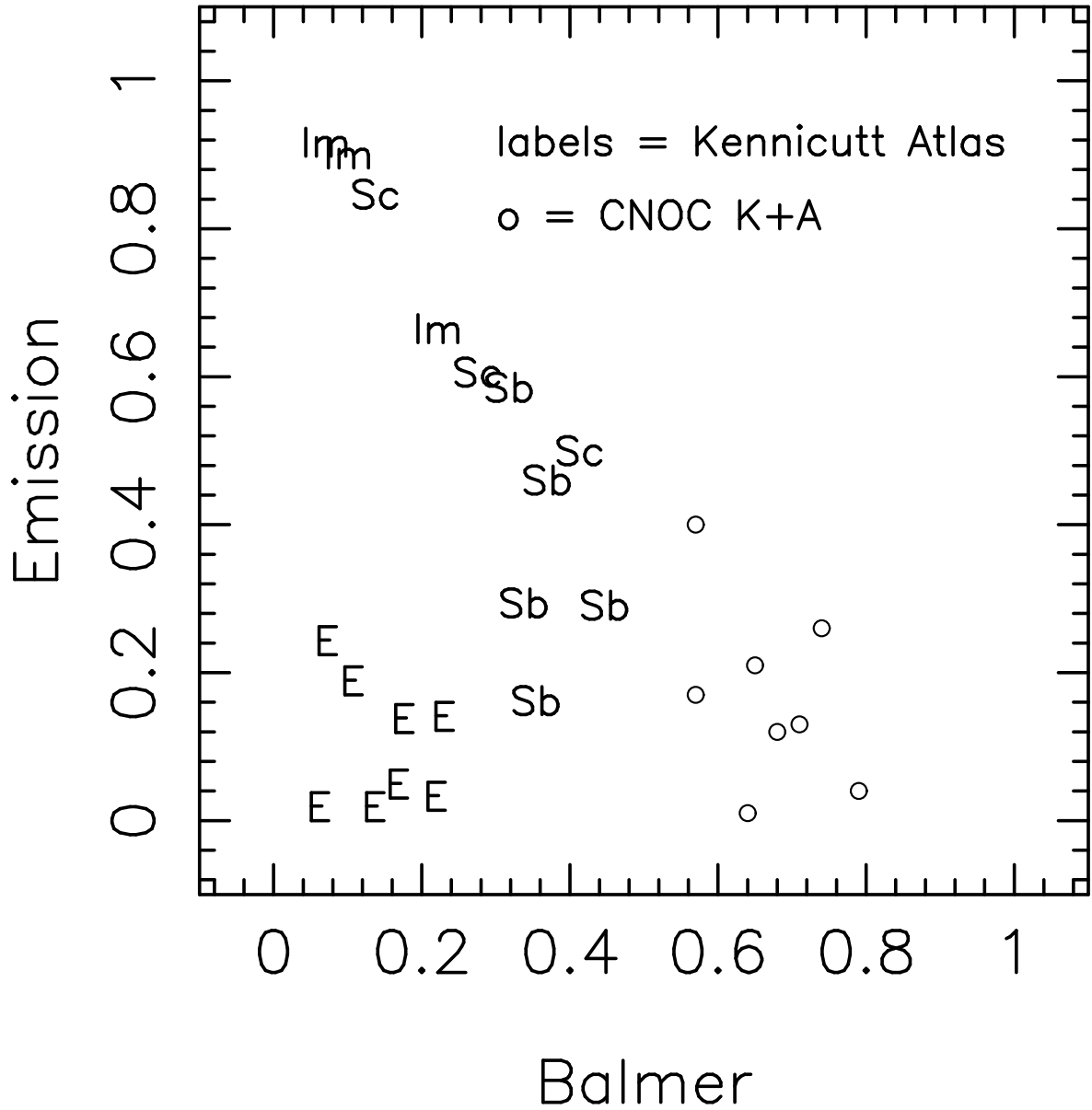}
\caption{PCA Decomposition of Galaxies from the Kennicutt 
Spectrophotometric Atlas (1992). Morphological types of the low redshift galaxies are 
labeled. Open circles denote K$+$A galaxies from the CNOC1 clusters.
The normalization of the sum of the three
components to 1 means that non-zero PCA components occupy a triangular region
on this plot. }
\end{figure}

\begin{figure}
\plottwo{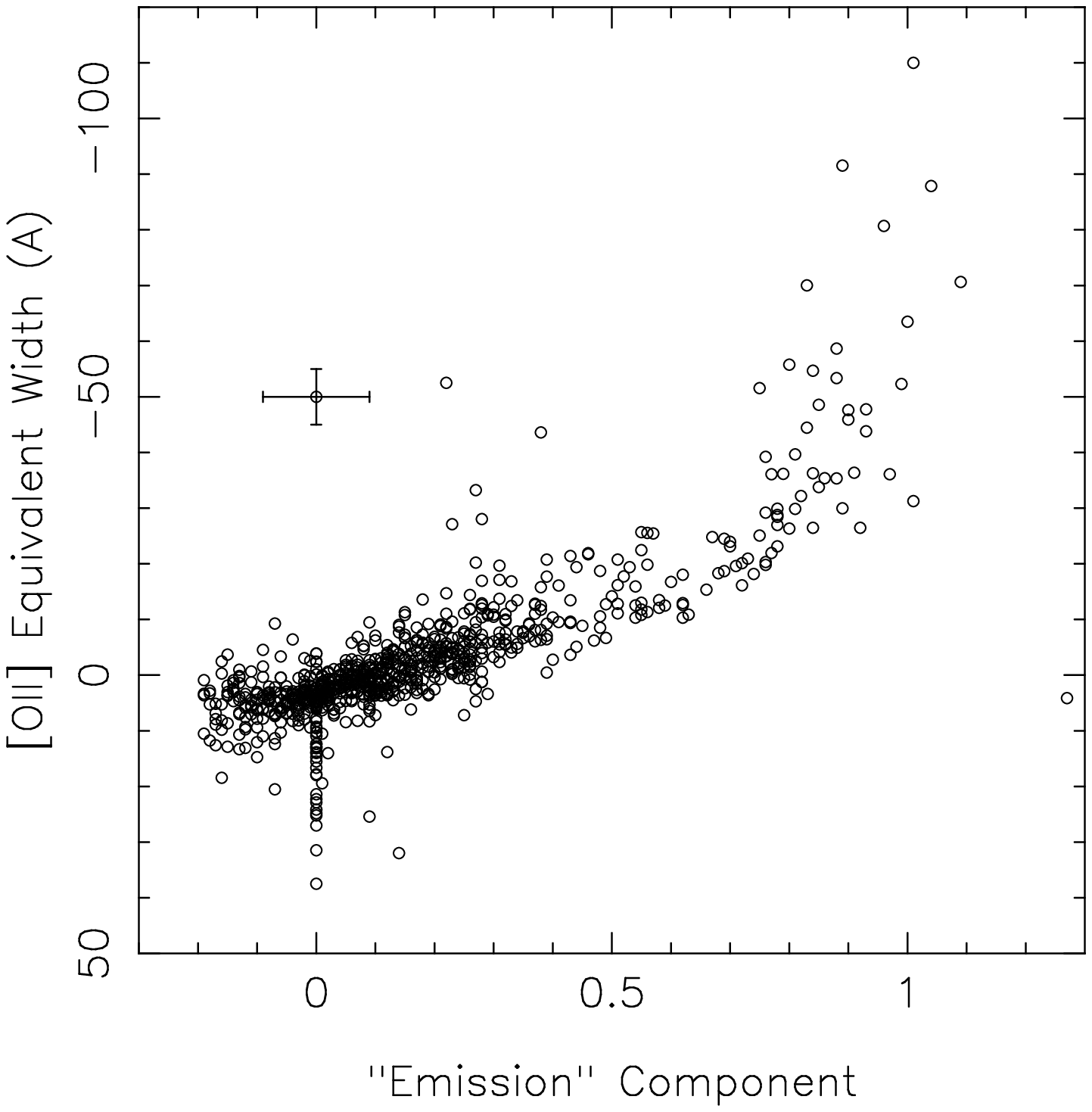}{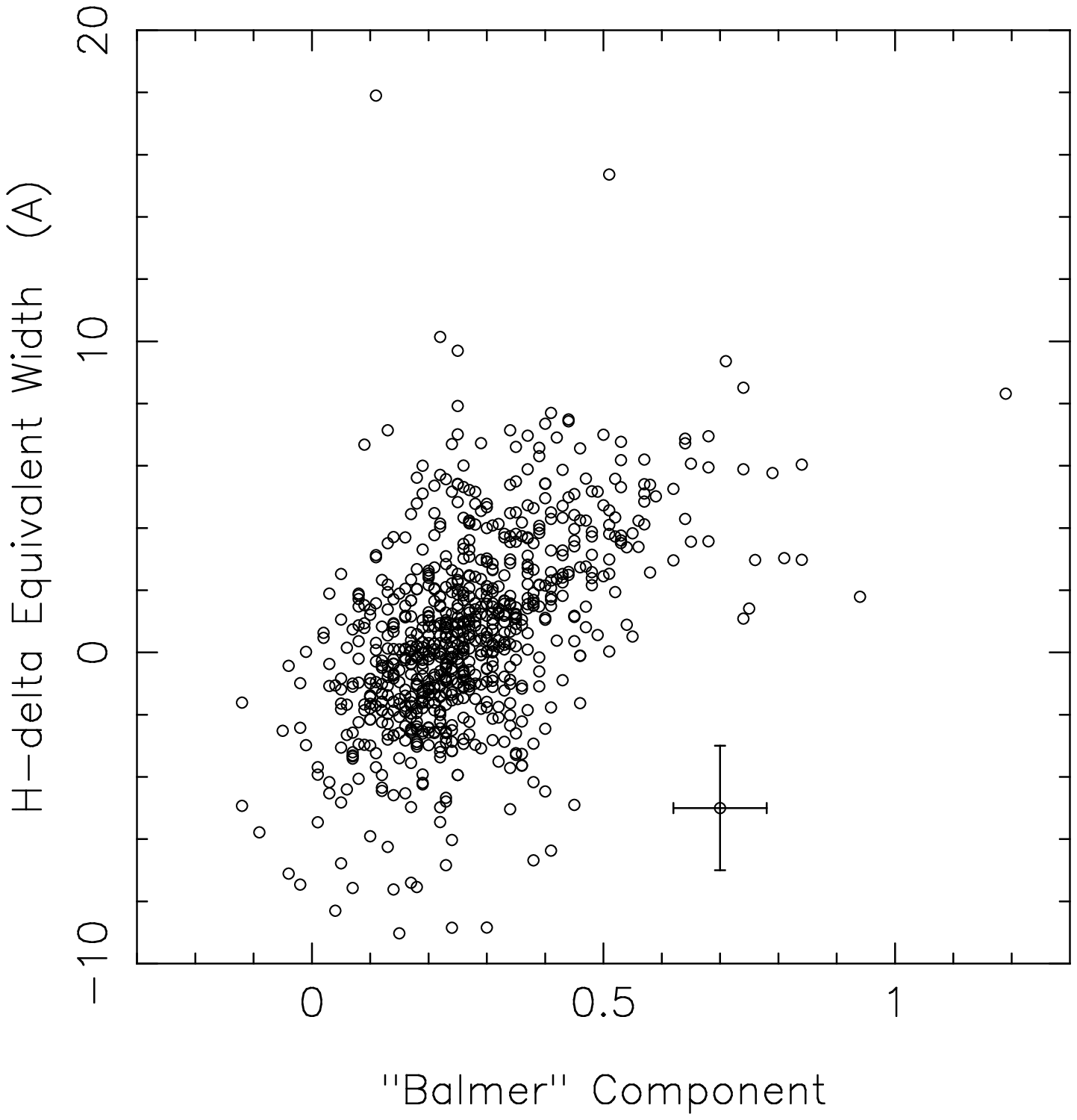}
\caption{PCA components versus line indices from B99.
Also plotted are median errorbars for the two measurements.
The expected  correlations between [OII] and ``Emission" and
H-$\delta$ and ``Balmer" are seen with a scatter
consistent with the observational error bars. Note that these correlations
are not required to pass through the origin, as the PCA components
are based on the relative strengths of many different lines.
}
\end{figure}

\begin{figure}
%\plotfiddle{allpop.ps}{3.0 truein}{0.0}{50}{50}{-200}{-50}
%\plotone{allpop.ps}
\plottwo{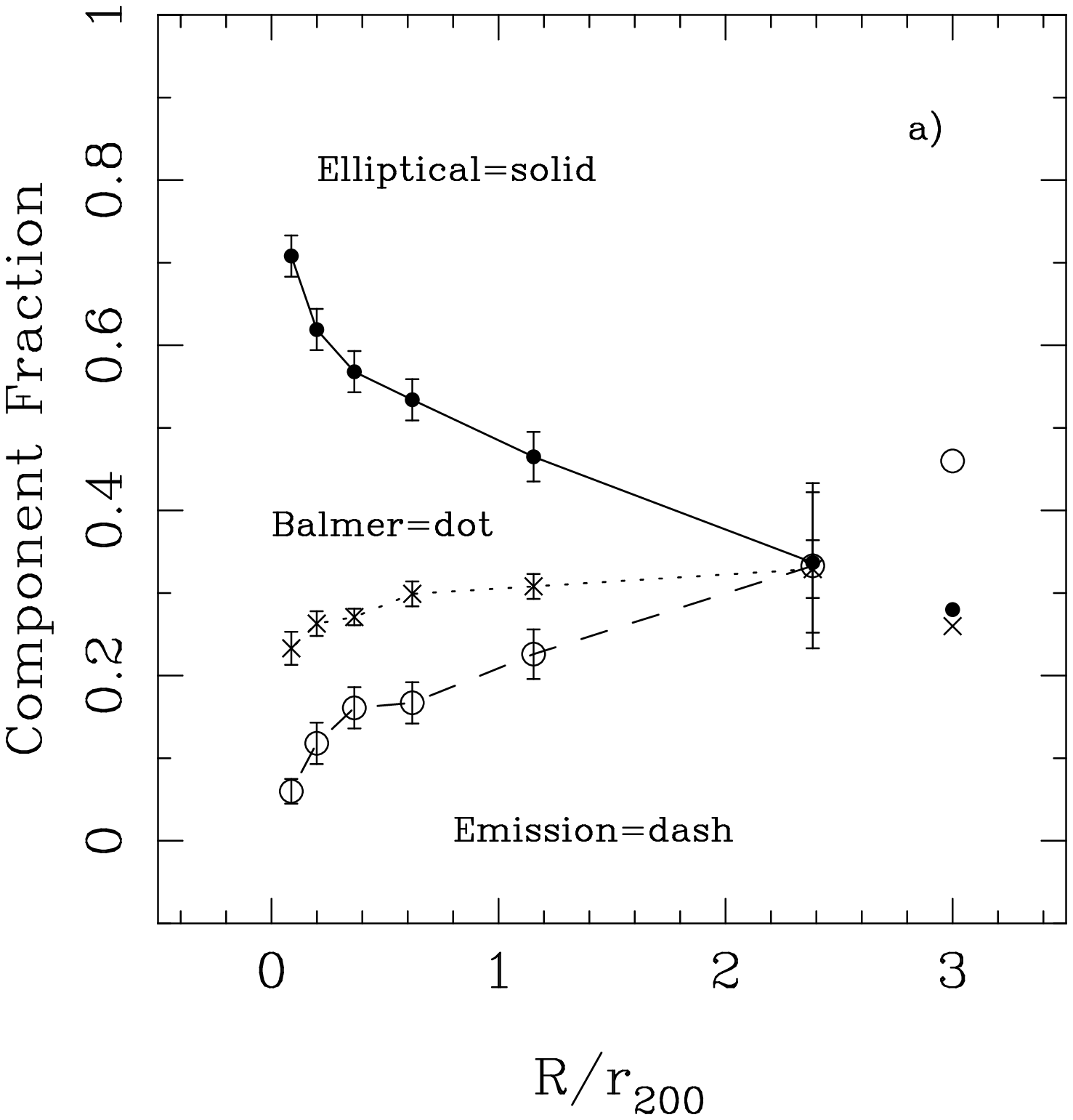}{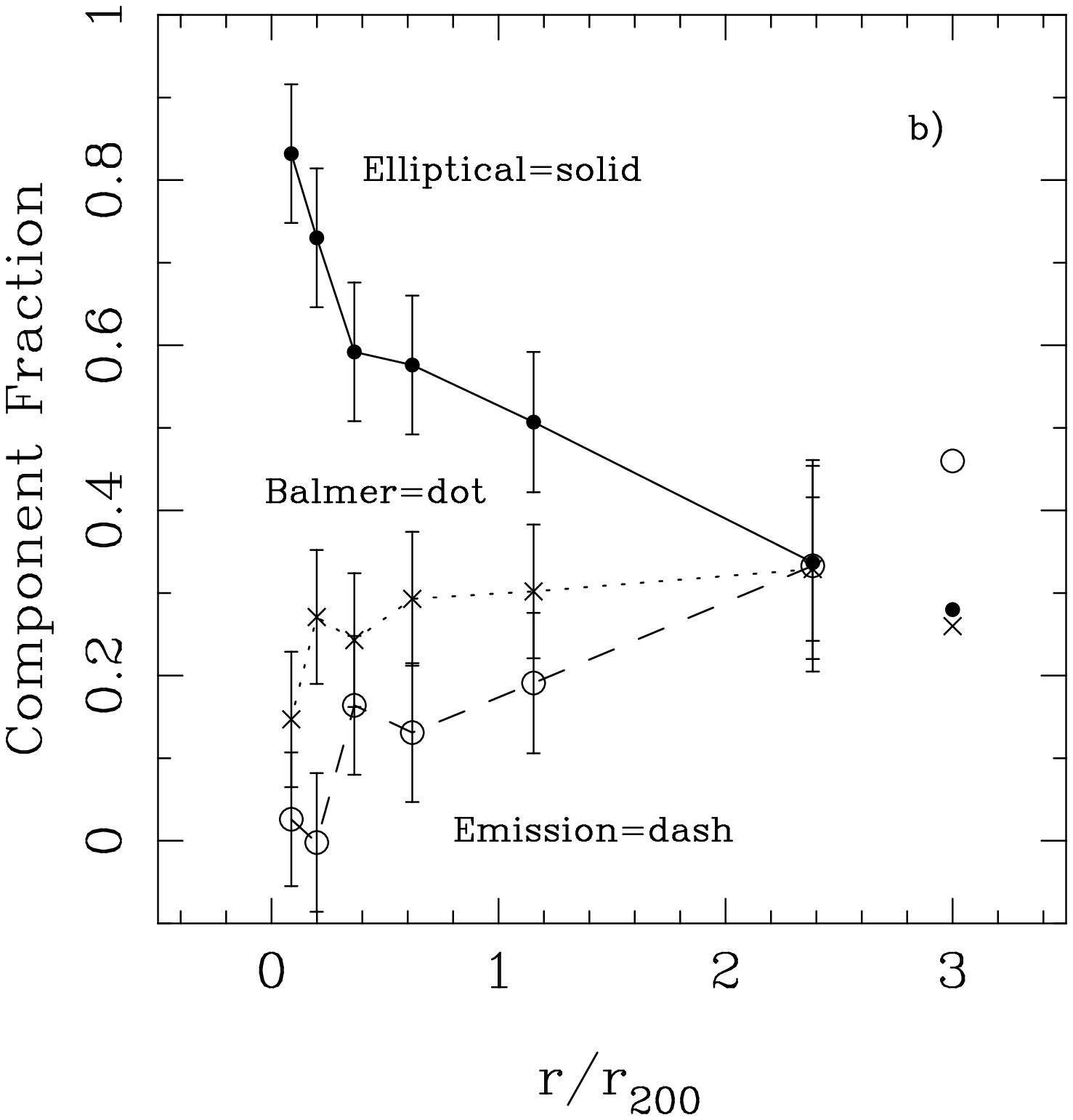}
\caption{Composite radial gradients from 913 galaxies in 15 clusters based on the
vectors shown in Figure 2.
The ``Elliptical" fraction is denoted
by closed circles and a solid line, the ``Emission" component by
open circles and a dashed line, and the ``Balmer" component by crosses and a dotted line.
The points at right represent average
values for field galaxies at $z \sim 0.35$ from the CNOC2
field galaxy survey. 
Panel a) is plotted in
projected radial coordinates, and panel b) in
deprojected radial coordinates (see text). }
\end{figure}

\begin{figure}
%\plotfiddle{allxfm.ps}{3.0 truein}{0.0}{50}{50}{-200}{-50}
\plottwo{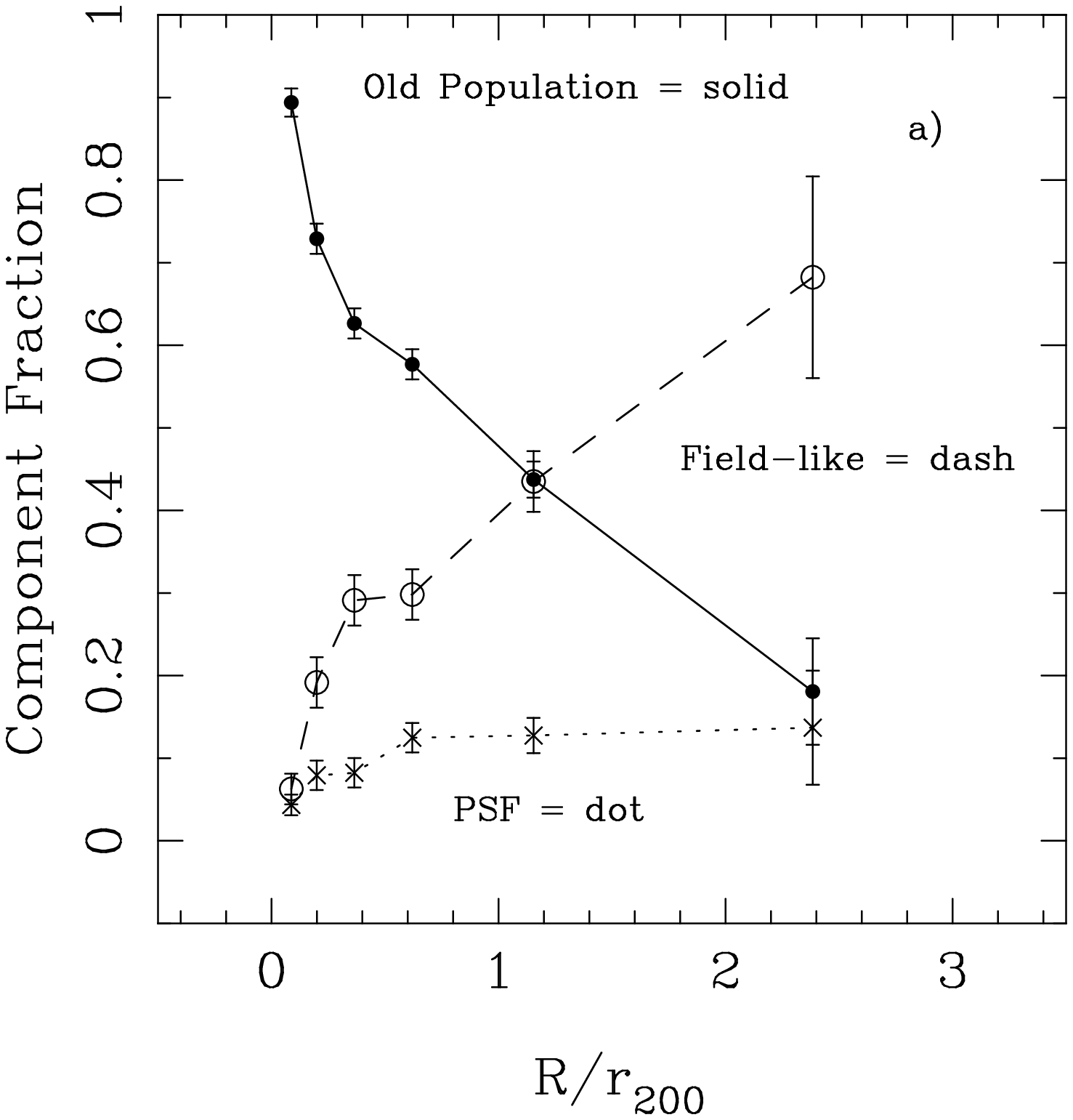}{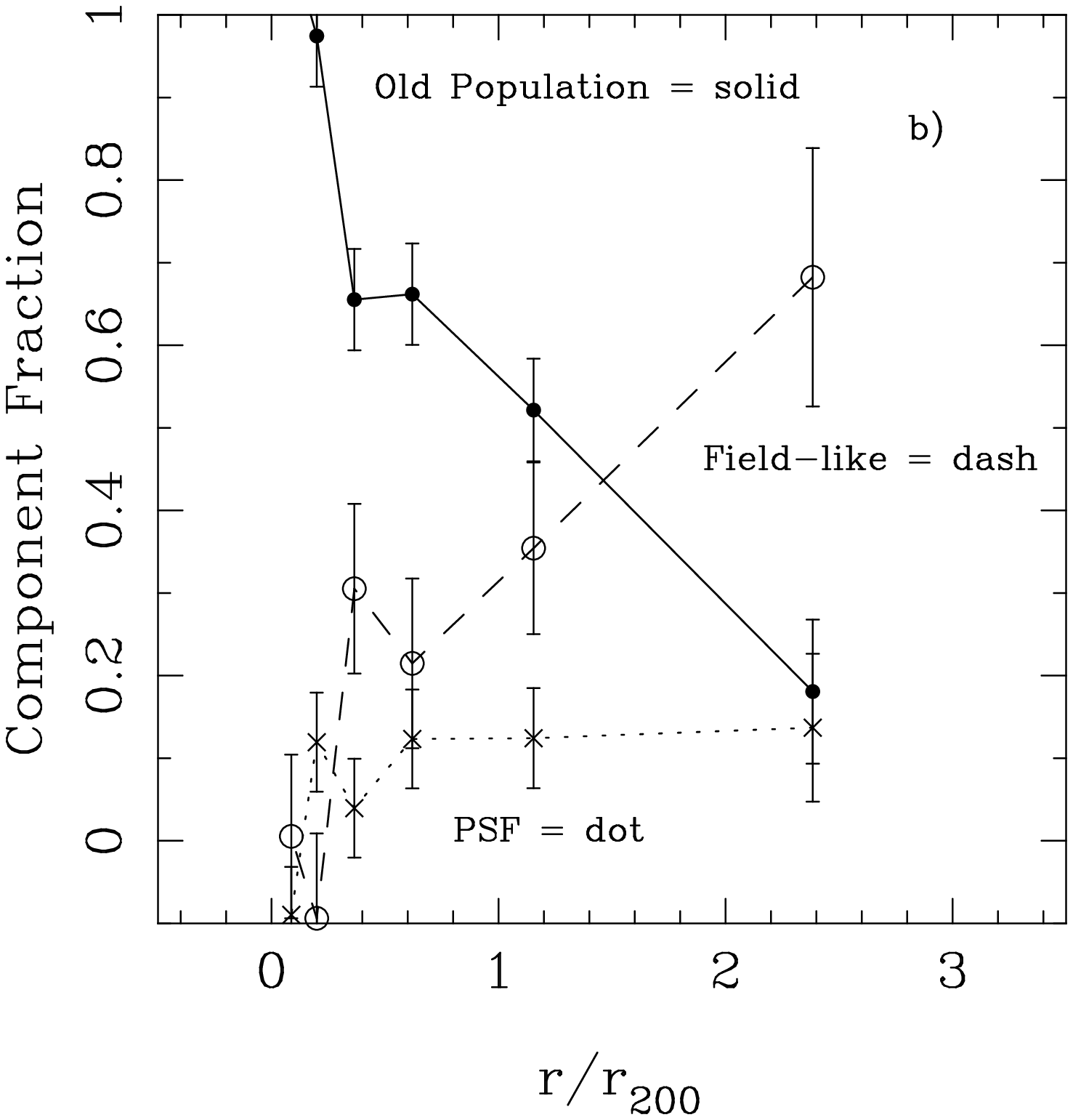}
\caption{ Composite radial gradients in transformed PCA coordinates
(see text). 
The ``Old Population" fraction is denoted
by closed circles and a solid line, the ``Field-like" component by
open circles and a dashed line, and the ``PSF" component by crosses and a dotted line.
Panel a) is plotted in
projected radial coordinates, and panel b) in
deprojected radial coordinates. }
\end{figure}

\begin{figure}
%\plotfiddle{zhilograd.ps}{3.0 truein}{0.0}{50}{50}{-200}{-50}
\plotone{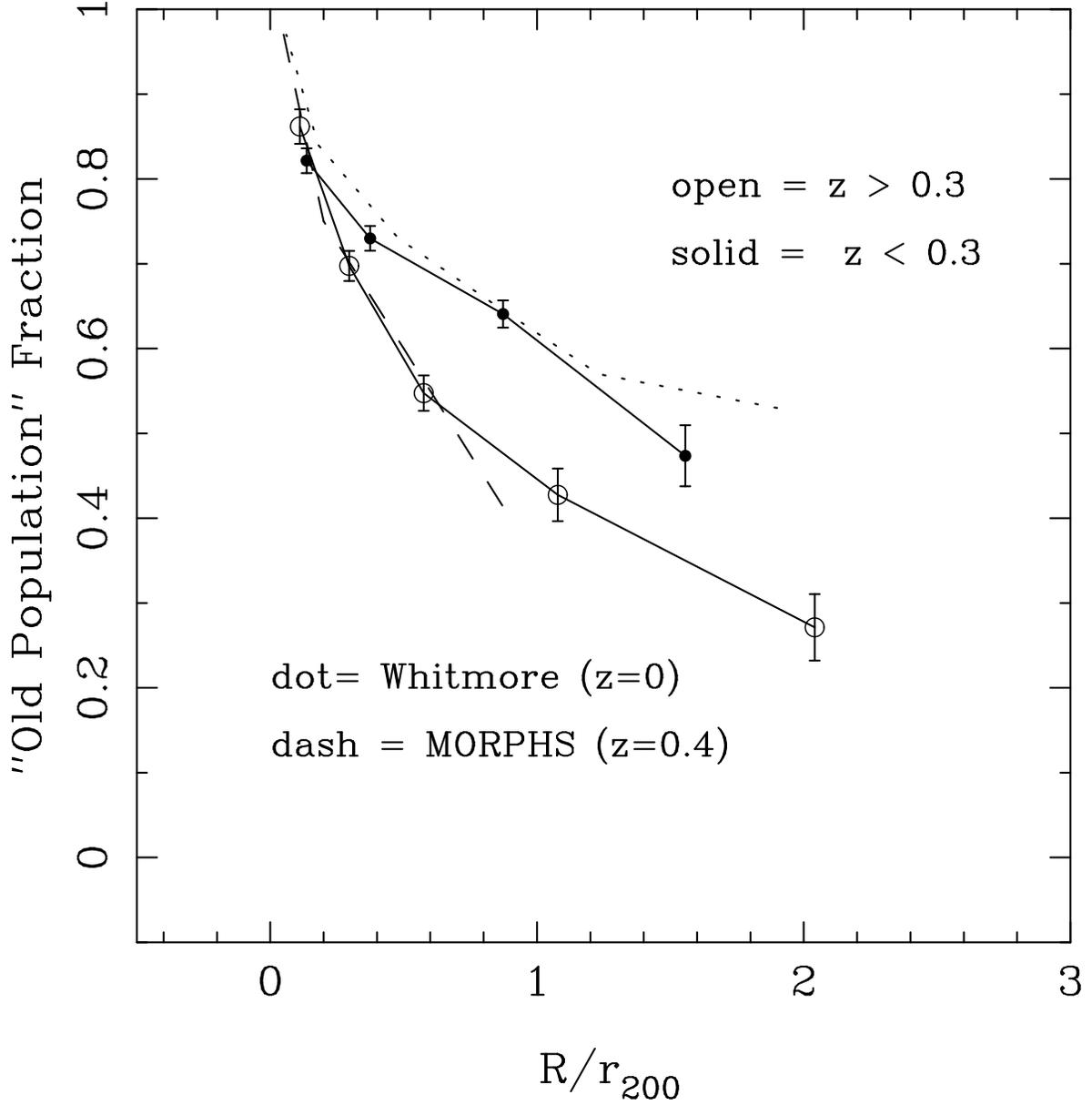}
\caption{ Composite radial gradients in the ``Old Population" component
for two redshift bins. Open and closed circles are
for CNOC1 cluster galaxies at $z > 0.3$, and $z < 0.3$, respectively. The dashed line 
represents the morphological gradients from
Dressler \etal (1997) for the $z \sim 0.4$ MORPHS sample,  and the dotted line 
is from gradients adapted from Whitmore \etal (1993)
at $z \sim 0$.}
\end{figure}

\begin{figure}
%\plotfiddle{allden.ps}{3.0 truein}{0.0}{50}{50}{-150}{-10}
\plotone{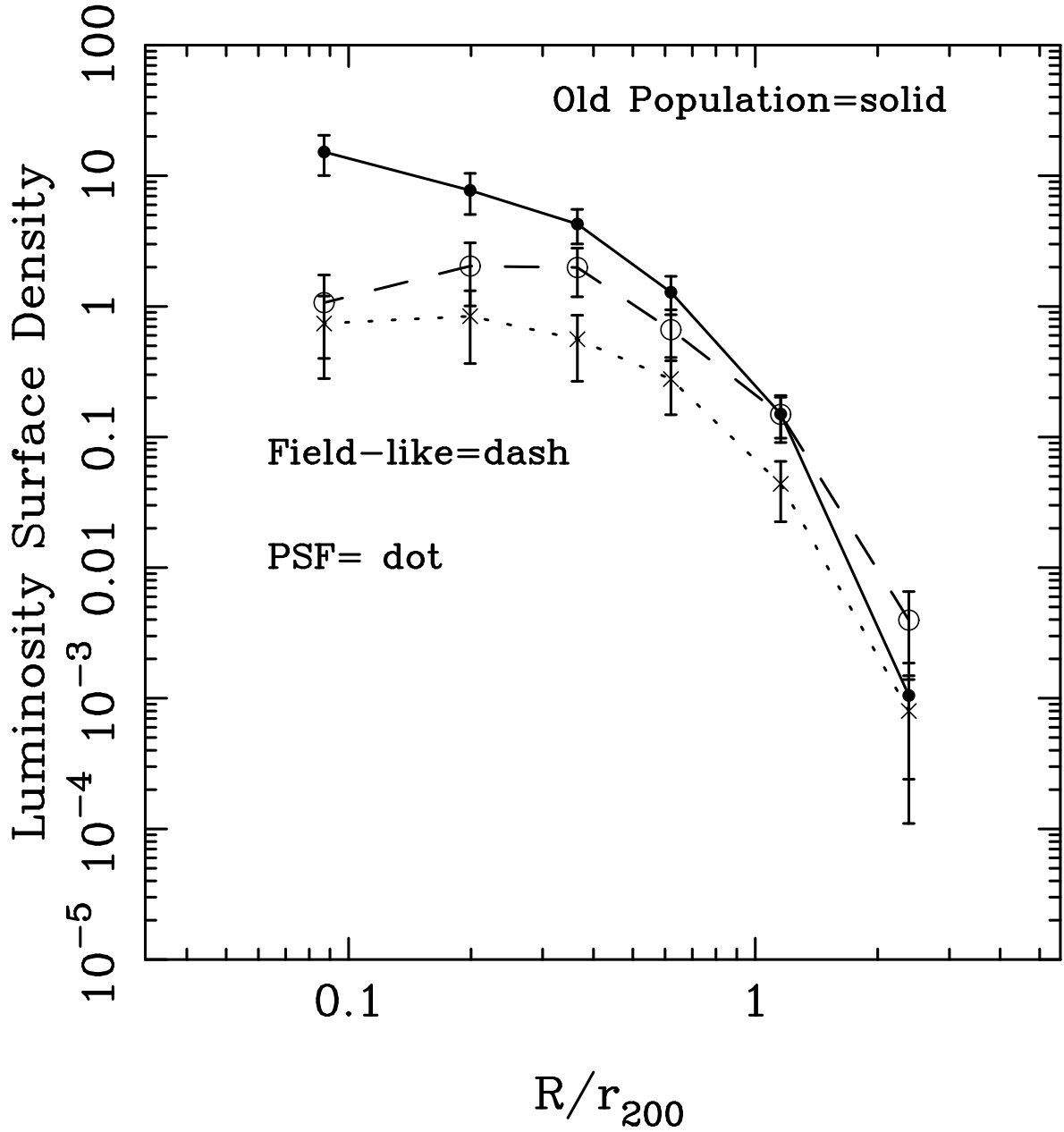}
\caption{ Composite luminosity density profiles from 15 clusters, in
relative units,
based on the transformed PCA coordinates.
The ``Old Population" fraction is denoted
by closed circles and a solid line, the ``Field-like" component by
open circles and a dashed line, and the ``PSF" component by crosses and a dotted line.
The profiles show
an increasing spatial extent from ``Old Population" to ``Post-Star-Formation"
to ``Field-like."}
\end{figure}

\begin{figure}
%\plotfiddle{zhilofden.ps}{1.0 truein}{0.0}{40}{40}{-25}{-3}
\plottwo{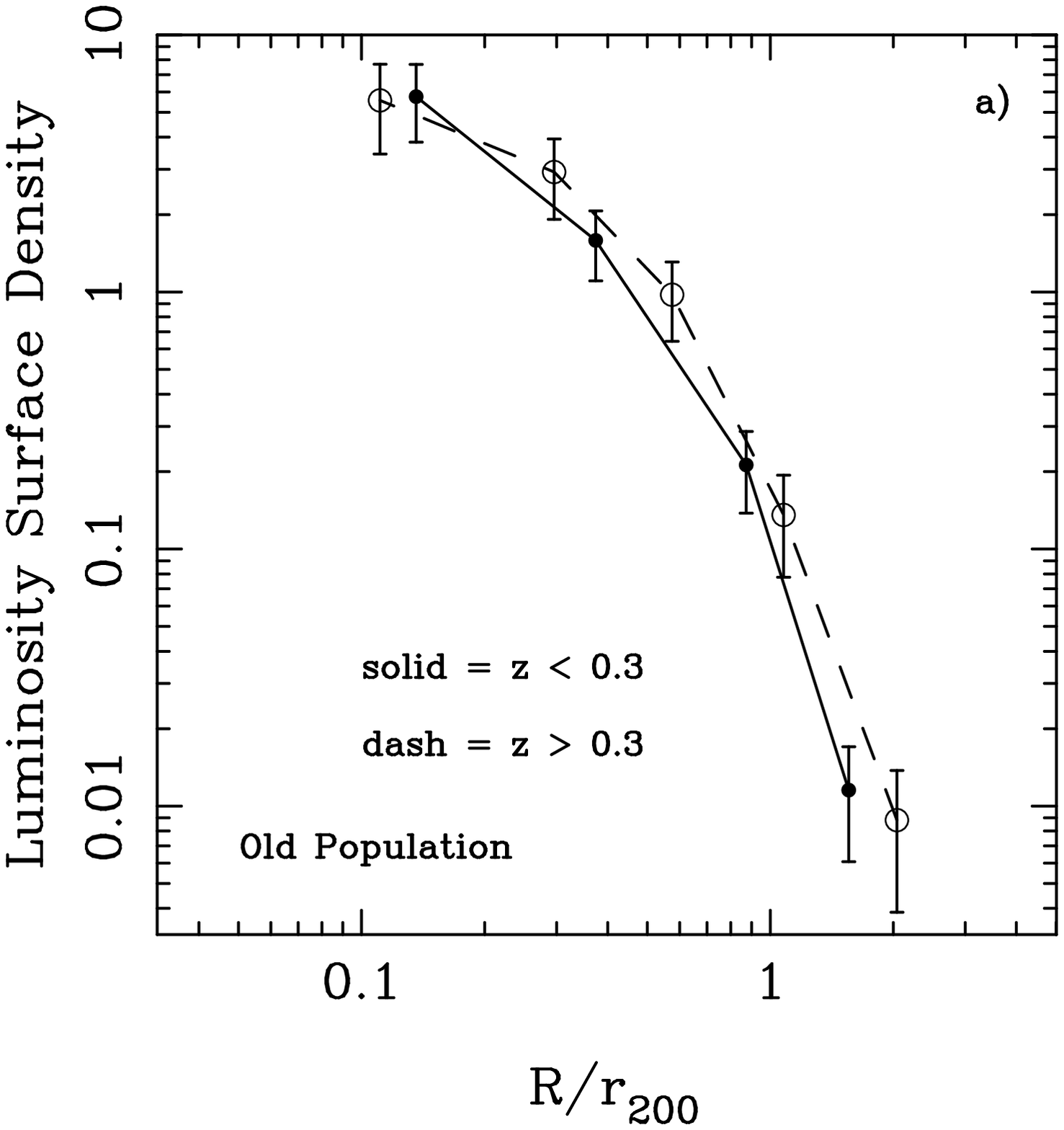}{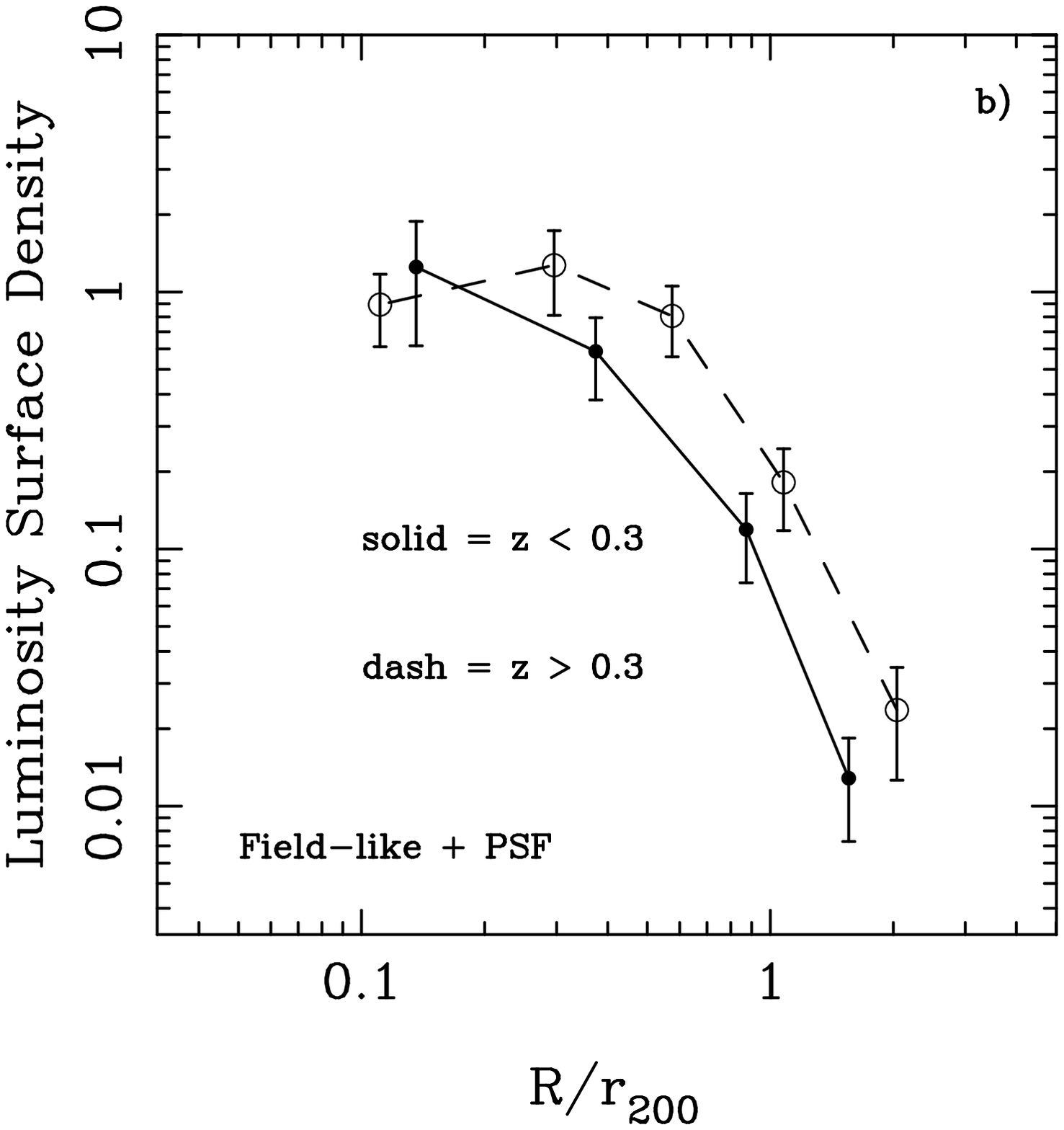}
%\plottwo{z1.ps}{z2.ps}
\caption{ Composite luminosity density profiles from 15 clusters,
in relative units,
for two subsamples at different redshifts.
Panel a)  shows profiles for the ``Old Population" component for
clusters with $0.18 < z < 0.30$ (solid  circles and line) and $0.30 < z < 0.55$ 
(open circles and dashed line).
Panel b) shows the
profiles for the sum of the ``Field-like" and  ``PSF" components and
the same redshift bins.}
\end{figure}

\begin{figure}
%\plotfiddle{xray.ps}{3.0 truein}{0.0}{50}{50}{-180}{-50}
\plotone{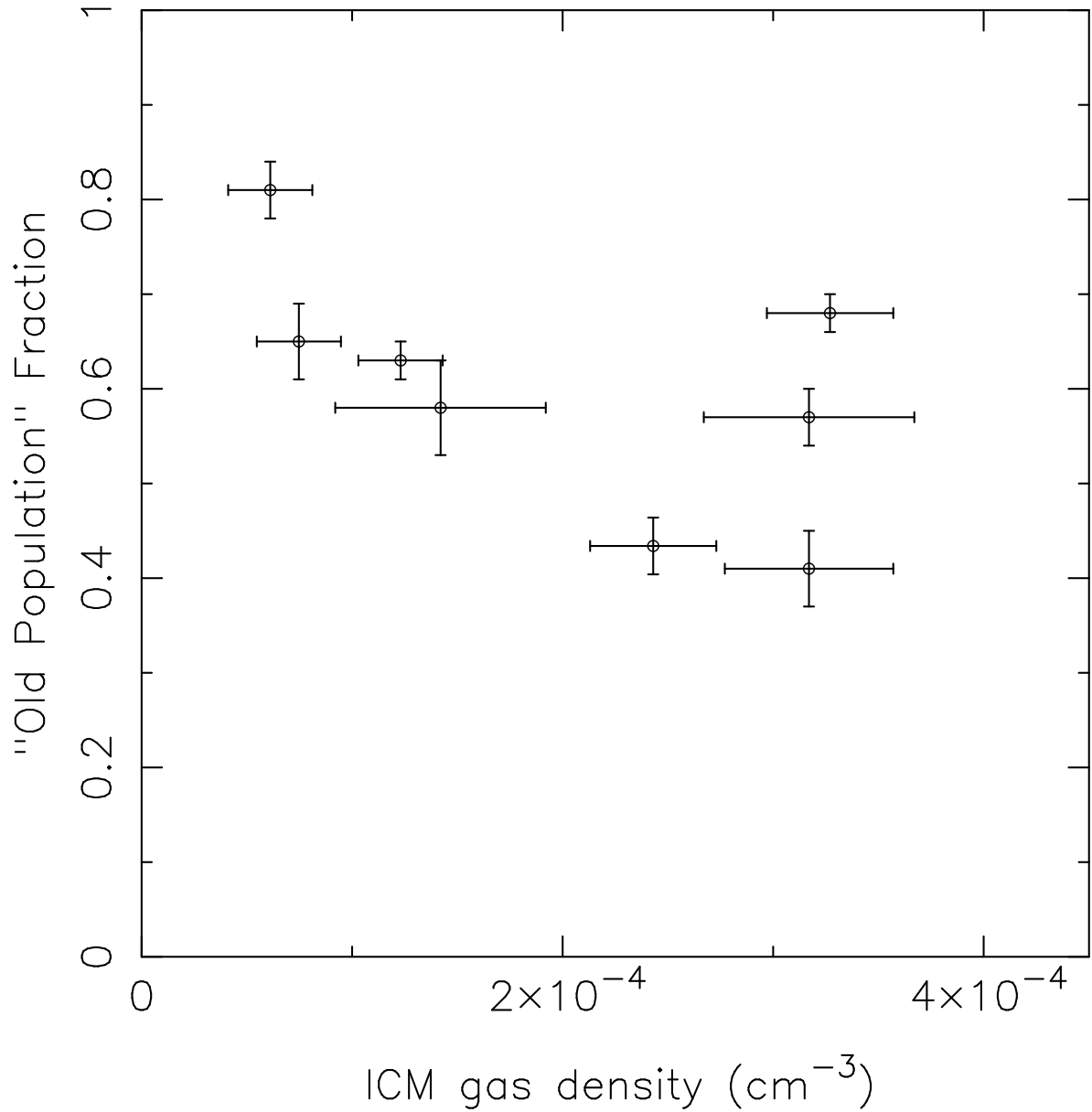}
\caption{ ``Old Population" component value versus X-ray gas density,
(units of cm$^{-3}$) as measured at 0.5 $r_{200}$.
 A direct link between cluster population evolution and X-ray
gas evolution would predict a positive correlation.
None is seen, suggesting that other factors drive the
observed evolution. } 
\end{figure}

\end{document}